\documentclass[onefignum,onetabnum]{siamonline220329}

\usepackage{xcolor}

\usepackage{amsmath, bm}
\usepackage{algorithm} 
\usepackage{algpseudocode}
\DeclareMathOperator\erf{erf}
\usepackage{multirow}
\usepackage{float}
\restylefloat{table}
\usepackage{mathpazo}
\usepackage{circuitikz}
\usetikzlibrary{patterns,hobby,decorations.pathmorphing}
\usepackage{tikz}
\usepackage{listofitems} % for \readlist to create arrays
\usetikzlibrary{shapes,arrows}
\usepackage{caption}
\usepackage{subcaption}
\usepackage{microtype}
\usepackage{lipsum}
\usepackage{amsfonts}
\usepackage{graphicx}
\usepackage{epstopdf}
\ifpdf
  \DeclareGraphicsExtensions{.eps,.pdf,.png,.jpg}
\else
  \DeclareGraphicsExtensions{.eps}
\fi

\usepackage{enumitem}
\setlist[enumerate]{leftmargin=.5in}
\setlist[itemize]{leftmargin=.5in}

\newsiamremark{remark}{Remark}
\newsiamremark{hypothesis}{Hypothesis}
\crefname{hypothesis}{Hypothesis}{Hypotheses}
\newsiamthm{claim}{Claim}

\usepackage{hyperref}
\usepackage{cleveref}

\headers{ACCRUE for Under and Overpredictions}{R. Bandy, E. Camporeale, A. Hu, T. Berger, and R. Morrison,}

\title{Accurate and Reliable Uncertainty Estimates for Deterministic Predictions: Extensions to Under and Overpredictions\thanks{Submitted to the editors April 8, 2026.
\funding{This work was funded by NASA under Grant No. 80NSSC20K1580 (Ensemble Learning for Accurate and Reliable Uncertainty Quantification).}}}

\author{Rileigh Bandy\thanks{Optimization and Uncertainty Quantification Department, Sandia National Laboratories, Albuquerque, NM
(\email{rjbandy@sandia.gov}).}
\and Enrico Camporeale\thanks{School of Physical and Chemical Sciences, Queen Mary University of London, London, UK and Space Weather Technology, Research and Education Center, University of Colorado, Boulder, CO
  (\email{Enrico.Camporeale@colorado.edu}).}
\and Andong Hu\thanks{Space Weather Technology, Research and Education Center, University of Colorado, Boulder, CO
  (\email{Andong.Hu@colorado.edu}).}
\and Thomas Berger \thanks{Mauna Loa Solar Observatory, National Center for Atmospheric Research, Boulder, CO
    (\email{tberger@ucar.edu }).}
\and Rebecca Morrison\thanks{Department of Computer Science, University of Colorado Boulder, Boulder, CO
(\email{rebeccam@colorado.edu}).}}

\usepackage{amsopn}

\ifpdf
\hypersetup{
  pdftitle={manuscript},
  pdfauthor={R. Bandy, A. Hu, T. Berger, E. Camporeale, and R. Morrison}
}
\fi

\begin{document}

\maketitle

% REQUIRED
\begin{abstract}
Computational models support high-stakes decisions across engineering and science, and practitioners increasingly seek probabilistic predictions to quantify uncertainty in such models. Existing approaches generate predictions either by sampling input parameter distributions or by augmenting deterministic outputs with uncertainty representations, including distribution-free and distributional methods. However, sampling-based methods are often computationally prohibitive for real-time applications, and many existing uncertainty representations either ignore input dependence or rely on restrictive Gaussian assumptions that fail to capture asymmetry and heavy-tailed behavior. Therefore, we extend the ACCurate and Reliable Uncertainty Estimate (ACCRUE) framework to learn input-dependent, non-Gaussian uncertainty distributions, specifically two-piece Gaussian and asymmetric Laplace forms, using a neural network trained with a loss function that balances predictive accuracy and reliability. Through synthetic and real-world experiments, we show that the proposed approach captures an input-dependent uncertainty structure and improves probabilistic forecasts relative to existing methods, while maintaining flexibility to model skewed and non-Gaussian errors.
\end{abstract}

% REQUIRED
\begin{keywords}
input-dependent uncertainty,
probabilistic prediction,
accuracy–reliability tradeoff,
skewed distributions
\end{keywords}

% REQUIRED
\begin{MSCcodes}
65C20, 62F25, 68T07, 62P12 
\end{MSCcodes}

\section{Introduction}
Computational models are often treated as black-boxes, where we know the inputs and outputs, but we do not have access to the precise mapping between the two. For these black-box models, outputs are usually deterministic, providing a single-point prediction. However, decision-making requires uncertainty quantification (UQ), motivating the development of \textit{a-posteriori} methods that augment deterministic model outputs with probabilistic predictions.

Classical approaches propagate uncertainty through input parameter distributions (e.g., ensemble or Bayesian sampling \cite{issan2023bayesian, luengo2020survey, sagi2018ensemble}). More generally, Bayesian frameworks have been developed to model both input and model-form uncertainty \cite{morrison2018representing, portone2022bayesian}, including flexible nonparametric approaches that capture complex features such as skewness and multimodality \cite{xie2021nonparametric}. Recent developments extend these ideas to neural network models, where uncertainty is represented through posterior distributions using structured priors and advanced sampling strategies \cite{chada2025bayesian}.
However, these approaches typically require posterior inference and repeated simulation, making them computationally prohibitive for expensive or high-dimensional models.

To address these challenges, deterministic outputs can instead be augmented with probabilistic uncertainty representations. These approaches broadly fall into distribution-free and distributional categories. Distribution-free approaches, such as conformal prediction (CP), construct prediction regions around model outputs with minimal assumptions \cite{taquet2022mapie}. Distribution-based approaches learn predictive distributions, either nonparametrically—via conditional cumulative distribution functions (CDFs), quantile estimation \cite{li2013optimal}, or regression-based methods such as EasyUQ \cite{walz2024easy}—or parametrically. While distribution-free and nonparametric methods offer flexibility, they can be difficult to interpret or deploy in structured settings. Parametric distribution-based approaches address these limitations by learning input-dependent uncertainty distributions;
in particular, the ACCurate and Reliable Uncertainty Estimate (ACCRUE) framework \cite{camporeale2021accrue, camporeale2019generation} learns a mapping from inputs to the parameters of a predictive distribution using a loss that balances accuracy and reliability.

Existing ACCRUE formulations assume Gaussian uncertainty, which limits their ability to capture skewness and heavy tails commonly observed in practice. In this work, we extend the ACCRUE framework to non-Gaussian uncertainty distributions, which allows for asymmetric and heavy-tailed predictive models. Specifically, we introduce two-piece Gaussian and asymmetric Laplace distributions within the ACCRUE loss framework in order to represent systematic biases and non-symmetric error structures. Our approach learns a catch-all uncertainty representation for observation–prediction pairs as a function of model inputs using a neural network parameterization. The resulting probabilistic predictions retain the interpretability and input dependence of ACCRUE while significantly increasing distributional flexibility.
% Uncertainties arise from many potential sources. Observation uncertainties commonly result from measurement errors, and model prediction uncertainties propagate from parameter errors and/or model-form errors. These errors tend to accumulate, and a catch-all uncertainty representation is formed when errors are not addressed in isolation. In general, we learn catch-all uncertainty representations for observation-prediction pairs as a function of the independent, input parameters of a black-box model. Specifically, we employ a neural network with a unique loss function to learn the parameters of the uncertainty distribution depending on the input parameters. The loss function balances an accuracy score with a reliability score and has been named the ACCurate and Reliable Uncertainty Estimate (ACCRUE) \cite{camporeale2019generation, camporeale2021accrue}. In ACCRUE the accuracy score measures how well on average an uncertain prediction matches an observation, while the reliability score discourages overfitting by measuring how well the aggregate empirical distribution of uncertainties matches the expected distribution.

The rest of the paper is organized as follows. In \Cref{sec:background}, we review ACCRUE with Gaussian distributed uncertainties. In \Cref{sec:extend}, we detail the methodology extending ACCRUE to skewed uncertainty distributions, specifically two-piece Gaussian and asymmetric Laplace distributions. We present synthetic results in \Cref{synthetic_section}, apply ACCRUE to weather forecasting in \Cref{sec:weather}, and give concluding remarks in \Cref{sec:conclusions}.

\section{Background: ACCRUE with Gaussian Distributed Uncertainties}
\label{sec:background}
In general, uncertainty in model predictions and observations are not uniformly distributed but instead functions of the inputs. Camporeale et al.\@~\cite{camporeale2019generation, camporeale2021accrue} developed a framework for adding a Gaussian uncertainty distribution with unique variance to each prediction. The uncertainty distribution is centered at the prediction $m$, its variance is dependent on inputs $\boldsymbol{X}$, and the error between an observation $y$ and the model prediction is $\varepsilon^o = y - m$. Therefore, the uncertainty distribution is defined as
\begin{equation}
    \mathcal{N}\left(m, \sigma(\boldsymbol{X})^2\right),
\end{equation}
and variance is learned. Specifically, variance is the output of a neural network (NN), where the inputs of the NN are the model inputs that generate the predictions. The functional relationship between the inputs and outputs is learned by minimizing the ACCRUE loss function.

ACCRUE balances accuracy and reliability, and it is defined as
\begin{equation}\label{accrue_eq}
    \text{ACCRUE} = \beta \cdot \overline{\text{CRPS}} + (1- \beta) \cdot \text{RS},
\end{equation}
where the mean continuous rank probability score ($\overline{\text{CRPS}}$) measures the average accuracy of the learned uncertainty distribution for all observation-prediction pairs. Reliability is measured by the reliability score ($\text{RS}$), which captures the mismatch between the aggregate empirical CDF and the aggregate expected CDF of the uncertainty for all observation-prediction pairs. Since accuracy and reliability tend to be competing objectives,
ACCRUE balances them. The choice of the weighting parameter $\beta$ introduces a tradeoff between competing objectives and can be interpreted as a hyperparameter tuning or bilevel optimization problem, where model parameters and calibration criteria are jointly balanced \cite{martin2025online}.

The accuracy term in ACCRUE for $N$ observation-prediction pairs with error\\ $\bm{\varepsilon}^o = \left(\varepsilon^o_1, \varepsilon^o_2, \hdots, \varepsilon^o_N \right)$ is defined by
\begin{equation}\label{general_crps}
    \overline{\text{CRPS}} = \frac{1}{N}\sum_{k=1}^N \int_{-\infty}^{\infty}
    \left[ P_k(\varepsilon) - H(\varepsilon - \varepsilon^o_k) \right]^2 d\varepsilon,
\end{equation}
where
\begin{equation}
    H(x) = 
    \begin{cases}
    0, \quad x< 0 \\
    1, \quad x \ge 0
    \end{cases}
\end{equation}
is the Heaviside step function that jumps from zero to one at the point where $\varepsilon$ equals the error $\varepsilon^o_k$, and $P_k(\varepsilon)$ is the CDF of the error between observation and prediction. Since we assume the uncertainty is normally distributed, 
\begin{equation}\label{normal_cdf}
    P_k(\varepsilon) = \frac{1}{2}\left[1+ \text{erf}\left(\frac{\varepsilon}{\sqrt{2}\sigma(\boldsymbol{X}_k)} \right) \right],
\end{equation}
where $\boldsymbol{X}_k$ are the inputs associated with the error $\varepsilon^o_k$. Substituting \Cref{normal_cdf} into \Cref{general_crps} yields
\begin{equation}\label{normal_crps}
    \overline{\text{CRPS}} = \frac{1}{N}\sum_{k=1}^N \sigma(\boldsymbol{X}_k)\left[\frac{\varepsilon^o_k}{\sigma(\boldsymbol{X}_k)}\text{erf}\left(\frac{\varepsilon^o_k}{\sqrt{2}\sigma(\boldsymbol{X}_k)}\right) + \sqrt{\frac{2}{\pi}}\exp\left(-\frac{{\left(\varepsilon^o_k\right)}^2}{2\sigma(\boldsymbol{X}_k)^2} \right) -\frac{1}{\sqrt{\pi}}\right].
\end{equation}

The reliability term in ACCRUE measures predictions' statistical consistency with observations, and it is defined by
\begin{equation}\label{gen_rs}
\text{RS} = \int_{-\infty}^{\infty} \left[
\Phi(\eta) - C(\eta)
\right]^2 d \eta,
\end{equation}
where $\eta$ is the standardized error, $\Phi(\eta)$ is the expected sample CDF, and $C(\eta)$ is the empirical CDF. Therefore, 
\begin{equation}\label{empirical_cdf}
    C(\eta) = \frac{1}{N}\sum_{i=1}^N H\left(\eta - \eta_i \right),
\end{equation}
where $\eta_i = \frac{\varepsilon^o_i}{\sqrt{2}\sigma(\boldsymbol{X}_i)}$. The standardized errors are assumed to be sorted $\{\eta_1, \eta_2, \hdots, \eta_N \}$ where $(\eta_i \le \eta_{i+1})$. Since we assume the uncertainty is normally distributed, 
\begin{equation}\label{norm_eta_cdf}
    \Phi(\eta) = \frac{1}{2}\left[1+ \text{erf}\left(\eta \right) \right].
\end{equation}
Substituting \Cref{norm_eta_cdf} into \Cref{gen_rs} yields
\begin{equation}\label{normal_rs}
    \begin{aligned}
\text{RS} &= 
\int_{-\infty}^{\infty} \left[
\frac{1}{2} + \frac{1}{2}\text{erf}\left(\eta \right)
 - \frac{1}{N} \sum_{i=1}^N H(\eta - \eta_i)
\right]^2 d \eta\\
&= 
\sum_{i=1}^N
\left[ \frac{\eta_i}{N}\left(\text{erf}(\eta_i) + 1 \right)
-  \frac{\eta_i}{N^2}\left(2i - 1\right)
+  \frac{\exp(-\eta_i^2)}{\sqrt{\pi} N}
\right] - \frac{1}{2}\sqrt{\frac{2}{\pi}}.
\end{aligned}
\end{equation}
% In the future, we will investigate how breaking the Gaussian distribution assumption of $\varepsilon^o_i$ impacts the RS. We hypothesize it will make the RS harder to minimize and a less effective regularizer.

To weigh the accuracy and reliability terms roughly equally---even if they have different magnitudes---we use the heuristic
\begin{equation}\label{beta_gauss}
    \beta = \frac{\text{RS}_{\text{min}}}{\overline{\text{CRPS}}_{\text{min}}
+\text{RS}_{\text{min}}}.
\end{equation}
The minimum of $\overline{\text{CRPS}}$ is
\begin{equation}
    \overline{\text{CRPS}}_{\text{min}} = \frac{\sqrt{\log4}}{2N}\sum_{i=1}^N \varepsilon_i,
\end{equation}
and the minimum RS is
\begin{equation}
    \text{RS}_{\text{min}} = \frac{1}{\sqrt{\pi}N}\sum_{i=1}^N\exp\left(-\left[\text{erf}^{-1}\left(\frac{2i-1}{N}-1 \right) \right]^2 \right) - \frac{1}{2}\sqrt{\frac{2}{\pi}}.
\end{equation}

\section{Extending ACCRUE}
\label{sec:extend}
It is a strong assumption that all uncertainties are normally distributed; this may fail under complex parameter uncertainty or measurement error, and especially in the face of significant model-form error.
In such cases, errors may be asymmetric for some inputs.
% and a natural extension from the Gaussian distribution to a skewed distribution is the log-normal distribution. We previously implemented ACCRUE for log-normal distributions \cite{bandy2023skewed} \red{reference a poster or no?}, but this has several shortcomings. First, the support is nonnegative. This requires all of the errors to be one-sided or a transformation of the errors (e.g., taking the absolute value of the errors or shifting them). Second, the log-normal distribution is always skewed in one direction, which is defined before learning the distribution. If data is left-skewed in one region and right-skewed in another, we have to manually classify the data before learning log-normal distribution parameters with ACCRUE. Finally, the log-normal distribution cannot be symmetric. Examples often exhibit regions of symmetric and asymmetric errors. While the log-normal distribution works in some cases, we opt for more robust distributions. 
We focus on distributions that can be left-skewed, right-skewed, or symmetric with a support of all real numbers since errors can generally be positive and negative. With those constraints in mind, we extend ACCRUE to other distributional forms---distributions with analytical CRPS and CDF solutions, specifically two-piece Gaussian distributions and asymmetric Laplace distributions. Analytical solutions to $\overline{\text{CRPS}}$ and RS are vital for efficient optimization of the ACCRUE loss function because numerically solving these integrals can be computationally expensive. Additionally, we leverage derivative information when optimizing ACCRUE, and automatic differentiation for numerical integration algorithms poses challenges that are adverted with analytical solutions \cite{eberhard1999automatic}. To learn an uncertainty distribution using ACCRUE, we must define the $\overline{\text{CRPS}}$, the RS, and the weighting coefficient $\beta$.

\textbf{Mean Continuous Rank Probability Score:} The general $\overline{\text{CRPS}}$ is
\begin{equation*}
\begin{aligned}
    \overline{\text{CRPS}} &= \frac{1}{N}\sum_{k=1}^N \text{CRPS}_k\\
    &= \frac{1}{N}\sum_{k=1}^N \int_{-\infty}^{\infty}
    \left[ P_k(\varepsilon) - H(\varepsilon - \varepsilon^o_k) \right]^2 d\varepsilon,
\end{aligned}
\end{equation*}
and we assume $\text{CRPS}_k$ has an analytical solution. There are numerous distributions with analytical CRPS solutions, and many are detailed by Jordan et al. \cite{jordan2017evaluating}. The main idea is to first evaluate the Heaviside function, which splits the integral into two pieces:
\begin{equation}
    \text{CRPS}_k = \int_{-\infty}^{\varepsilon^o_k}
    \left[ P_k(\varepsilon) - 0 \right]^2 d\varepsilon + \int^{\infty}_{\varepsilon^o_k}
    \left[ P_k(\varepsilon) - 1 \right]^2 d\varepsilon.
\end{equation}
Next, the analytical CDF is substituted into $P_k(\varepsilon)$. Depending on the distribution, deriving the analytical CRPS can be simple or tedious, but the approach is straightforward. We detail the analytical CRPS solutions for the two-piece Guassian distribution in \Cref{tpg} and the asymmetric Laplace distribution in \Cref{al}.

\textbf{Reliability Score:} The general RS is
\begin{equation*}
    \text{RS} = \int_{-\infty}^{\infty} \left[ \Phi(u) - C(u) \right]^2 du,
\end{equation*}
where the aggregate empirical CDF is the average Heaviside function for each transformed error $C(u) = \frac{1}{N}\sum_{i=1}^N H\left(u - u_i \right)$. The aggregate expected CDF is standard uniform $\Phi(u) = u$. A CDF transform (probability integral transform) is done to convert the expected distribution to uniform, and transformed errors are assumed to be sorted $\{u_1, u_2, \hdots, u_N \}$ where $(u_i \le u_{i+1})$. Then, the analytical solution to the RS with a standard uniform aggregate expected CDF is

\begin{subequations}\label{general_rs}
    \begin{align}
        \text{RS} &= \int_{-\infty}^{\infty} \left[ \Phi(u) - C(u) \right]^2 du = \int_{-\infty}^{\infty} \left[ \Phi(u) - \frac{1}{N}\sum_{i=1}^N H\left(u - u_i \right) \right]^2 du \label{rs_def} \\
        &= \int_{0}^{1} \left[ u - \frac{1}{N}\sum_{i=1}^N H\left(u - u_i \right) \right]^2 du \label{rs_uniform} \\
        &= \int_{0}^{1} u^2 d u
        \quad - 2\frac{1}{N} \int_{0}^{1} u \sum_{i=1}^N H\left(u - u_i \right) d u 
        \quad + \frac{1}{N^2} \int_{0}^{1} \left[\sum_{i=1}^N H\left(u - u_i \right) \right]^2 d u \label{rs_split} \\
    &= \frac{1}{3} - u_N + \frac{1}{N}\sum_{i=1}^N u_i^2 + \frac{1}{N^2}\sum_{i=1}^{N-1} i^2 \left(u_{i+1} - u_i \right). \label{rs_soln}
    \end{align}
\end{subequations}
By definition, we have \Cref{rs_def}. Then the standard uniform CDF is substituted in for $\Phi(u)$ in \Cref{rs_uniform}, and the integral bounds are changed to $u \in [0,1]$ to match the CDF's support. In \Cref{rs_split}, the integral is expanded and split into three parts. The final analytical solution is given in \Cref{rs_soln}.

This RS is more general than the original, Gaussian RS defined in \Cref{normal_rs}, but we still assume an expected uncertainty distribution for the CDF transform. More work is needed to investigate how the RS behaves when the errors are not distributed as assumed. We hypothesize it will make the RS harder to minimize and a less effective regularizer, but prior knowledge can be leveraged to select the expected uncertainty distribution and hopefully mitigate the problem.

\textbf{$\bm \beta$ Heuristic:} 
Now that we have analytical solutions to the two pieces of ACCRUE---the accuracy piece $\overline{\text{CRPS}}$ and the reliability piece RS---we need to combine them. We follow Camporeale's original formulation:
\begin{equation*}
    \text{ACCRUE} = \beta \cdot \overline{\text{CRPS}} + (1- \beta) \cdot \text{RS},
\end{equation*}
where we constrain $\beta \in (0,1)$. With $\beta=0$ the loss function is informed solely by the RS, which could result in underfitting. With $\beta=1$ the loss function becomes $\overline{\text{CRPS}}$, which could result in overfitting \cite{camporeale2021accrue}.

% RB: should change this to algorithmic package instead
\begin{algorithm}[ht!]
\caption{Generalized $\beta$ heuristic.}
\begin{algorithmic}[1]
\Function{$\beta$\_search}{$\bm \varepsilon^o$}
\State $\bm \beta \gets [0.1, 0.2, \hdots, 0.9]$
\State $dist = \infty$
\State $\beta^* = -1$
\For{$\beta_i$ in $\bm \beta$}
\State $CRPS_i, RS_i \gets \beta\_CALIBRATION(\beta_i, \bm \varepsilon^o)$
\State $dist_i \gets \sqrt{CRPS_i^2 + RS_i^2}$
\If{$dist_i < dist$}
\State $dist \gets dist_i$
\State $\beta^* \gets \beta_i$
\EndIf
\EndFor
\State \Return $\beta^*$
\EndFunction
\end{algorithmic}
\label{beta_search}
\end{algorithm}

In this generalized format, we often optimize for more than one parameter of an uncertainty distribution. This complicates minimizing $\overline{\text{CRPS}}$ and RS: With respect to which parameter do we differentiate? Additionally, while the CRPS must have an analytical solution, real roots to the CRPS's partial derivatives are not required. Therefore, we replace \Cref{beta_gauss} with a generalized approach for selecting $\beta$ as detailed in \Cref{beta_search}. A grid search of $\beta$-values from $0.1$ to $0.9$ is performed on training data $\bm \varepsilon^o$. For a specific $\beta_i$, the NN is calibrated on the training data, and the resulting $\overline{\text{CRPS}}$ and RS values for the training data are returned. Then, the $L^2$ norm of $\overline{\text{CRPS}}$ and RS is computed, and the $\beta$-value that minimized this distance $\beta^*$ is returned as the optimal weighting between accuracy and reliability.

\subsection{ACCRUE with Two-Piece Gaussian Uncertainties}\label{tpg}

We extend ACCRUE to the two-piece Gaussian distribution, a continuous distribution consisting of two Gaussian distributions of potentially unequal scales with its mode at $\varepsilon^o_k$ and adjusted to assure continuity and normalization.

The PDF of the uncertainty at $\varepsilon^o_k$ with inputs $\bm X_k$ is
\begin{equation}
    p_k(\varepsilon) =
    \begin{cases}
        & \frac{\sqrt{2}}{\sqrt{\pi}(\sigma_1(\bm{X}_k) + \sigma_2(\bm{X}_k))} \exp\left(-\frac{\varepsilon^2}{2\sigma_1(\bm{X}_k)^2} \right) \quad \text{if}\ \varepsilon \le 0 \\
        & \frac{\sqrt{2}}{\sqrt{\pi}(\sigma_1(\bm{X}_k) + \sigma_2(\bm{X}_k))} \exp\left(-\frac{\varepsilon^2}{2\sigma_2(\bm{X}_k)^2} \right) \quad \text{if}\ \varepsilon > 0
    \end{cases},
\end{equation}
where $\sigma_1(\bm X_k ), \sigma_2(\bm X_k )  \in \mathbb{R}_{>0}$ are scale parameters. When $\sigma_1(\bm X_k ) = \sigma_2(\bm X_k )$, the two-piece Gaussian distribution becomes symmetric and corresponds to the Gaussian distribution. The CDF of the two-piece Gaussian distribution is
\begin{equation}\label{tpg_cdf}
\begin{aligned}
    P_k(\varepsilon) &=
    % \begin{cases}
    %     & \frac{2\sigma_1(\bm X_k )}{\sigma_1(\bm{X}_k) + \sigma_2(\bm{X}_k)} \Phi \left(\frac{\varepsilon}{\sigma_1(\bm X_k )} \right) \text{if}\ \varepsilon \le 0 \\
    %     & \frac{\sigma_1(\bm X_k ) - \sigma_2(\bm X_k )}{\sigma_1(\bm{X}_k) + \sigma_2(\bm{X}_k)} + \frac{2\sigma_2(\bm X_k )}{\sigma_1(\bm{X}_k) + \sigma_2(\bm{X}_k)} \Phi \left(\frac{\varepsilon}{\sigma_2(\bm X_k )} \right) \text{if}\ \varepsilon > 0
    % \end{cases} \\
    % &= 
    \begin{cases}
        & \frac{\sigma_1(\bm X_k )}{\sigma_1(\bm{X}_k) + \sigma_2(\bm{X}_k)} + \frac{\sigma_1(\bm X_k )}{\sigma_1(\bm{X}_k) + \sigma_2(\bm{X}_k)} \text{erf} \left(\frac{\varepsilon}{\sqrt{2}\sigma_1(\bm X_k )} \right) \quad \text{if}\ \varepsilon \le 0 \\
        & \frac{\sigma_1(\bm X_k )}{\sigma_1(\bm{X}_k) + \sigma_2(\bm{X}_k)} + \frac{\sigma_2(\bm X_k )}{\sigma_1(\bm{X}_k) + \sigma_2(\bm{X}_k)} \text{erf} \left(\frac{\varepsilon}{\sqrt{2}\sigma_2(\bm X_k )} \right) \quad \text{if}\ \varepsilon > 0
    \end{cases}.
\end{aligned}
\end{equation}
Then, the CDF is used to transform the errors $\bm \varepsilon^o$ into standard uniform errors $\bm \eta$. Specifically, $\varepsilon_k^o$ $\forall k \in (1, 2, \hdots, N)$ are input into \Cref{tpg_cdf}, and the resulting $\eta_k$ are sorted such that $\bm \eta = \{\eta_1, \eta_2, \hdots, \eta_N \}$ where $(\eta_i \le \eta_{i+1})$.

\textbf{Analytical CRPS for the two-piece Gaussian distribution:}
From Gneiting and Thorarinsdottir, the analytical solution to the two-piece Gaussian CRPS \cite{gneiting2010predicting} is
% \begin{equation}
% \begin{aligned}
% \overline{\text{CRPS}} &= \frac{1}{N}\sum_{k=1}^N \int_{-\infty}^{\infty}
%     \left[ P_k(\varepsilon) - H(\varepsilon - \varepsilon^o_k) \right]^2 d\varepsilon \\
% &= \frac{1}{N}\sum_{k=1}^N
% \begin{cases}
% \begin{aligned}
%     &\frac{4\sigma_1(\bm X_k )^2}{\sigma_1(\bm X_k )+\sigma_2(\bm X_k )}\left[\frac{\varepsilon^o_k}{\sigma_1(\bm X_k )} \Phi\left(\frac{\varepsilon^o_k}{\sigma_1(\bm X_k )}\right) + \phi\left(\frac{\varepsilon^o_k}{\sigma_1(\bm X_k )}\right)\right] \\
%     &- \varepsilon^o_k
%     + \frac{2}{\sqrt{\pi}} \frac{\sqrt{2}\sigma_2(\bm X_k ) \left(\sigma_2(\bm X_k )^2-\sigma^2_1\right) - \left(\sigma_1(\bm X_k )^3+\sigma_2(\bm X_k )^3\right)}{\left(\sigma_1(\bm X_k )+\sigma_2(\bm X_k )\right)^2}
% \end{aligned}
%  \quad \text{if } \varepsilon^o_k\le 0 \\
% \begin{aligned}
%      &\frac{4\sigma_2(\bm X_k )^2}{\sigma_1(\bm X_k )+\sigma_2(\bm X_k )}\left[\frac{\varepsilon^o_k}{\sigma_2(\bm X_k )} \Phi\left(\frac{\varepsilon^o_k}{\sigma_2(\bm X_k )}\right) + \phi\left(\frac{\varepsilon^o_k}{\sigma_2(\bm X_k )}\right)\right] \\
%     &+ \frac{\left(\sigma_1(\bm X_k )-\sigma_2(\bm X_k )\right)^2-4\sigma_2(\bm X_k )^2}{\left(\sigma_1(\bm X_k )+\sigma_2(\bm X_k )\right)^2}\varepsilon^o_k \\
%     &+ \frac{2}{\sqrt{\pi}} \frac{\sqrt{2}\sigma_1(\bm X_k ) \left(\sigma_1(\bm X_k )(\bm X_k )^2-\sigma^2_2\right) - \left(\sigma_1(\bm X_k )^3+\sigma_2(\bm X_k )^3\right)}{\left(\sigma_1(\bm X_k )+\sigma_2(\bm X_k )\right)^2} 
% \end{aligned}
% \quad \text{if } \varepsilon^o_k> 0
% \end{cases},
% \end{aligned}
% \end{equation}
\begin{equation}
\begin{aligned}
\overline{\text{CRPS}} &= \frac{1}{N}\sum_{k=1}^N \int_{-\infty}^{\infty}
    \left[ P_k(\varepsilon) - H(\varepsilon - \varepsilon^o_k) \right]^2 d\varepsilon \\
&= \frac{1}{N}\sum_{k=1}^N
\begin{cases}
\begin{aligned}
    &\frac{4\sigma_1^2}{\sigma_1+\sigma_2}\left[\frac{\varepsilon^o_k}{\sigma_1} \Phi\left(\frac{\varepsilon^o_k}{\sigma_1}\right) + \phi\left(\frac{\varepsilon^o_k}{\sigma_1}\right)\right] 
    % \\
    - \varepsilon^o_k
    \\ 
    &+ \frac{2}{\sqrt{\pi}} \frac{\sqrt{2}\sigma_2 \left(\sigma_2^2-\sigma^2_1\right) - \left(\sigma_1^3+\sigma_2^3\right)}{\left(\sigma_1+\sigma_2\right)^2}
\end{aligned}
 \qquad \qquad \qquad \qquad \text{ if } \varepsilon^o_k\le 0 \\
\begin{aligned}
     &\frac{4\sigma_2^2}{\sigma_1+\sigma_2}\left[\frac{\varepsilon^o_k}{\sigma_2} \Phi\left(\frac{\varepsilon^o_k}{\sigma_2}\right) + \phi\left(\frac{\varepsilon^o_k}{\sigma_2}\right)\right]
    + \frac{\left(\sigma_1-\sigma_2\right)^2-4\sigma_2^2}{\left(\sigma_1+\sigma_2\right)^2}\varepsilon^o_k \\
    &+ \frac{2}{\sqrt{\pi}} \frac{\sqrt{2}\sigma_1 \left(\sigma_1^2-\sigma^2_2\right) - \left(\sigma_1^3+\sigma_2^3\right)}{\left(\sigma_1+\sigma_2\right)^2} 
\end{aligned}
\quad \text{if } \varepsilon^o_k> 0
\end{cases},
\end{aligned}
\end{equation}
where $\phi(x) = \frac{1}{\sqrt{2\pi}}\exp\left(-\frac{x^2}{2}\right)$ is the standard normal PDF, and $\Phi(x) = \frac{1}{2}\left[ 1+ \erf (x)\right]$ is the standard normal CDF. Note, we shorten $\sigma_1 = \sigma_1(\bm X_k)$ and $\sigma_2 = \sigma_2(\bm X_k)$ for legibility.

\subsection{ACCRUE with Asymmetric Laplace Uncertainties}\label{al}
Next, we extend ACCRUE to the asymmetric Laplace distribution, a continuous distribution consisting of two exponential distributions of unequal scales back-to-back at $\varepsilon^o$ and adjusted to assure continuity and normalization.

The PDF of the uncertainty at $\varepsilon^o_k$ with inputs $\bm X_k$ is
\begin{equation}
    p_k(\varepsilon) =
    \begin{cases}
        & \frac{\lambda(\bm X_k)}{\kappa(\bm X_k) + 1/\lambda(\bm X_k)} \exp\left(\frac{\lambda(\bm X_k)}{\kappa(\bm X_k)} \varepsilon \right) \qquad \qquad \;  \text{ if } \varepsilon \le 0 \\
        & \frac{\lambda(\bm X_k)}{\kappa(\bm X_k) + 1/\lambda(\bm X_k)} \exp\left(-\lambda(\bm X_k)\kappa(\bm X_k) \varepsilon \right) \quad \text{if}\ \varepsilon > 0
    \end{cases},
\end{equation}
where $\lambda(\bm X_k) \in \mathbb{R}_{>0}$ is the scale parameter, and $\kappa(\bm X_k) \in \mathbb{R}_{>0}$ is the asymmetry parameter. When $\kappa(\bm X_k) = 1$, the asymmetric Laplace distribution becomes symmetric and corresponds to the Laplace distribution.
The CDF of the error between an observation-prediction pair becomes
\begin{equation}\label{al_cdf}
    P_k(\varepsilon) =
    \begin{cases}
        &\frac{\kappa(\bm X_k)^2}{1 + \kappa(\bm X_k)^2} \exp \left( \frac{\lambda(\bm X_k)}{\kappa (\bm X_k)}\varepsilon \right) 
        \qquad \qquad \qquad \;\text{if}\ \varepsilon \le 0 \\
        &1 -  \frac{1}{1 + \kappa(\bm X_k)^2} \exp \left(- \lambda(\bm X_k)\kappa (\bm X_k)\varepsilon \right) \quad \text{if}\ \varepsilon > 0
    \end{cases}.
\end{equation}
Then, the CDF is used to transform the errors $\bm \varepsilon^o$ into standard uniform errors $\bm \eta$. 

\textbf{Analytical CRPS for the asymmetric Laplace distribution:}
% If $\varepsilon^o_k \le 0:$
% \begin{equation}
% \begin{aligned}
% \overline{\text{CRPS}} &= \frac{1}{N}\sum_{k=1}^N \int_{-\infty}^{\infty}
%     \left[ P_k(\varepsilon) - H(\varepsilon - \varepsilon^o_k) \right]^2 d\varepsilon \\
% &= \frac{1}{N}\sum_{k=1}^N
% \begin{cases}
% \begin{aligned}
%     \left| \varepsilon_k^o \right|
% &+ \frac{2\kappa(\bm X_k)^3}{\lambda(\bm X_k) \left(1+\kappa(\bm X_k)^2\right)} \left[\exp \left(\frac{\lambda(\bm X_k)}{\kappa(\bm X_k)}\varepsilon_k^o \right) - 1\right]
% \\
% &+ \frac{\kappa(\bm X_k)^5}{2\lambda(\bm X_k) \left(1+\kappa(\bm X_k)^2\right)^2} +
% \frac{1}{2\lambda(\bm X_k) \kappa(\bm X_k) \left(1+\kappa(\bm X_k)^2\right)^2}
% \end{aligned}
% \qquad \text{if}\ \varepsilon_k^o \le 0 \\
% \begin{aligned}
%     \left| \varepsilon_k^o \right|
% &+ \frac{2}{\lambda(\bm X_k) \kappa(\bm X_k) \left(1+\kappa(\bm X_k)^2\right)} \left[\exp \left(-\lambda(\bm X_k)\kappa(\bm X_k)\varepsilon_k^o \right) -1 \right]
% \\
% &+ \frac{\kappa(\bm X_k)^5}{2\lambda(\bm X_k) \left(1+\kappa^2\right)^2}
% + \frac{1}{2\lambda(\bm X_k) \kappa(\bm X_k) \left(1+\kappa(\bm X_k)^2\right)^2} 
% \end{aligned}
% \qquad \text{if}\ \varepsilon_k^o > 0
% \end{cases}
% \end{aligned}    
% \end{equation}
\begin{equation}
\begin{aligned}
\overline{\text{CRPS}} &= \frac{1}{N}\sum_{k=1}^N \int_{-\infty}^{\infty}
    \left[ P_k(\varepsilon) - H(\varepsilon - \varepsilon^o_k) \right]^2 d\varepsilon \\
&= \frac{1}{N}\sum_{k=1}^N
\begin{cases}
\begin{aligned}
    \left| \varepsilon_k^o \right|
&+ \frac{2\kappa^3}{\lambda \left(1+\kappa^2\right)} \left[\exp \left(\frac{\lambda}{\kappa}\varepsilon_k^o \right) - 1\right]
\\
&+ \frac{\kappa^5}{2\lambda \left(1+\kappa^2\right)^2} +
\frac{1}{2\lambda \kappa \left(1+\kappa^2\right)^2}
\end{aligned}
\qquad \text{if}\ \varepsilon_k^o \le 0 \\
\begin{aligned}
    \left| \varepsilon_k^o \right|
&+ \frac{2}{\lambda \kappa \left(1+\kappa^2\right)} \left[\exp \left(-\lambda\kappa\varepsilon_k^o \right) -1 \right]
\\
&+ \frac{\kappa^5}{2\lambda \left(1+\kappa^2\right)^2}
+ \frac{1}{2\lambda \kappa \left(1+\kappa^2\right)^2} 
\end{aligned}
\qquad \text{if}\ \varepsilon_k^o > 0
\end{cases}
\end{aligned}    
\end{equation}
which is the analytical solution of the CRPS for the asymmetric Laplace (two-piece exponential) distribution from \cite{jordan2017evaluating} with a change of variables ($\sigma_1=\frac{\kappa}{\lambda}$ and $\sigma_2=\frac{1}{\lambda\kappa}$). Note, we shorten $\kappa = \kappa(\bm X_k)$ and $\lambda = \lambda(\bm X_k)$ for legibility.
\section{Synthetic Results}\label{synthetic_section}
When comparing real-world observations to model predictions, uncertainties manifest from many sources. Observations often include measurement errors, and predictions come with parameter errors, model-form errors, or a combination. Distinguishing and quantifying these errors is challenging, and black-box models further complicate those distinctions. 
% Instead of looking at a system of equations or a coupled physics model and using expert knowledge to illicit the error's source, the errors in the predictions are propagated to the output together making them almost impossible to detangle. 
In this section, we test whether ACCRUE can find the true underlying measurement error by comparing noisy observations to predictions without parameter or model-form errors. All code—to run ACCRUE, generate data, and postprocess—is available here: \href{https://github.com/rbandy/ACCRUE_for_skewed}{github.com/rbandy/ACCRUE\_for\_skewed} \cite{rileigh_bandy_2026_19343211}.

\subsection{Generating Synthetic Data} We take $N=10,000$ observation-prediction pairs resulting in vectors of length $N$ for the inputs $\bm x = \left(x_1, x_2, \hdots, x_N \right)$, observations \\$\bm y = \left(y_1, y_2, \hdots, y_N \right)$, predictions $\bm m = \left(m_1, m_2, \hdots, m_N \right)$, and errors $\bm \varepsilon^o = \left( \varepsilon^o_1, \varepsilon^o_2, \hdots, \varepsilon^o_N \right)$.

For a single observation-prediction pair, the input is $x_k \in [0,1] \ \forall k \in [1,N]$. The observation is
\begin{equation}
    y_k = 0 + \delta\left(\bm \theta(x_k)\right),
\end{equation}
where $\delta\left(\bm \theta(x_k)\right)$ is synthetic measurement error sampled from a distribution with input-dependent parameters $\bm \theta$. The prediction is the observation without any measurement error, $m_k = 0$.
The error between the observation and prediction is $\varepsilon^o_k = y_k - m_k$.

\textbf{Measurement Error:} We generate a few sets of synthetic data with measurement error from unique distributions. The measurement error is distributed as either a two-piece Gaussian or asymmetric Laplace. In both cases, the distribution parameters are modeled as functions of the input. In the two-piece Gaussian distribution the parameters are $\bm \theta(x_k) = \left(\theta_1(x_k), \theta_2(x_k) \right) = \left(\sigma_1(x_k), \sigma_2(x_k) \right)$, and in the asymmetric Laplace distribution  the parameters $\bm \theta(x_k) = \left(\theta_1(x_k),\theta_2(x_k) \right) =\left(\kappa(x_k),  \frac{1}{\lambda(x_k)} \right)$.

In real-world observations, these parameter functions could vary in complexity from constant to strongly nonlinear. For our synthetic examples, we consider linear and nonlinear, trigonometric functions. The linear parameter functions are
\begin{equation}
    \theta^{\text{lin}}_1(x_k)=0.5x_k + 0.5
\end{equation}
and
\begin{equation}
    \theta^{\text{lin}}_2(x_k)=-2x_k + 2.5.
\end{equation}
The trigonometric parameter functions are
\begin{equation}
    \theta^{\text{trig}}_1(x_k)=\exp\left(\sin(2\pi x_k)\right)/3
\end{equation}
and
\begin{equation}
    \theta^{\text{trig}}_2(x_k)=\cos(2\pi x_k)+2.
\end{equation}

\subsection{Calibration}\label{accrue_calibration}
Given synthetic inputs and errors we use an NN to represent the distribution parameters, and we train the NN by minimizing the ACCRUE loss function. The result is a parametric model of input-dependent distribution parameters $\bm \theta$. In the following calibration scenarios, we know the true parameter functions that generate the measurement error, so we directly compare them to the ones we learn with ACCRUE.

\textbf{Calibration Scenarios:}
For the two-piece Gaussian and asymmetric Laplace distributions, we try three parameter function combinations: 
\begin{equation*}
    \text{linear: } \bm \theta^{\text{lin}} = \left(\theta^{\text{lin}}_1(x_k), \theta^{\text{lin}}_2(x_k) \right),
\end{equation*}
\begin{equation*}
    \text{trigonometric: } \bm \theta^{\text{trig}} = \left(\theta^{\text{trig}}_1(x_k), \theta^{\text{trig}}_2(x_k) \right),
\end{equation*}
and
\begin{equation*}
    \text{combination: } \bm \theta^{\text{combo}} = \left(\theta^{\text{lin}}_1(x_k), \theta^{\text{trig}}_2(x_k) \right).
\end{equation*}
Altogether there are six calibration scenarios as shown in \Cref{synthetic_scenarios}, where TPG is the two-piece Gaussian distribution and AL is the asymmetric Laplace distribution.

\begin{table}[ht]
\centering
\begin{tabular}{|c|c|c|c|c|c|c|}
\hline
\textbf{Scenario}                                                      & A      & B             & C           & D  & E  & F  \\ \hline
\textbf{Distribution}                                                  & TPG & TPG & TPG& AL & AL & AL \\ \hline
\textbf{\begin{tabular}[c]{@{}c@{}}Parameter\\ Functions\end{tabular}} & Linear & Trig & Combo & Linear & Trig & Combo    \\ \hline
\end{tabular}
\caption{Synthetic data calibration scenarios}
\label{synthetic_scenarios}
\end{table}

For each scenario, a set of synthetic data is generated to set the $\beta$-value in \Cref{accrue_eq}. We employ the grid search heuristic from \Cref{beta_search}, where the calibration data is all $N=10,000$ errors and associated inputs. Within the calibration, the set is randomly split into $80\%$ training and $20\%$ validation data. Once the optimal $\beta$-value, $\beta^*$, is determined, it is held constant for subsequent parameter calibration. 

Next, we generate a $100$-member ensemble of trained NNs to decrease prediction error \cite{kotu2018data}. Each member is given a set of $N=10,000$ randomly generated synthetic data points. Each set is randomly split into $64\%$ training, $16\%$ validation, and $20\%$ testing data. For each ensemble member, an NN learns the parameter functions of the distribution to minimize ACCRUE. 
% The NN architecture is depicted in \Cref{nn_synthetic_data}. 
The input layer is one-dimensional since the synthetic inputs are one-dimensional. The input layer is fully connected to a hidden layer of 10 nodes, and the activation function is the rectified linear unit (ReLU). The hidden layer is fully connected to the output layer, and the activation function is leakyReLU. Finally, the output layer is two-dimensional, representing the distribution parameters. For the two-piece Gaussian and asymmetric Laplace distributions, the parameters must be nonnegative and greater than zero. We enforce this constraint by exponentiating the outputs of the NN. Additional NN hyperparameters are as follows: the optimizer is ADAM, the learning rate is $0.005$, the batch size is $100$, and learning is stopped at $1,000$ epochs or when the validation loss has not improved for the last $10$ epochs.

% \tikzstyle{mynode}=[thick,draw=gray,fill=gray!20,circle,minimum size=22]
% \begin{figure}[H]
%     \centering
%     \begin{tikzpicture}[x=2.5cm,y=1.0cm]
%       \readlist\Nnod{1,10,2} % number of nodes per layer
%       % \Nnodlen = length of \Nnod (i.e. total number of layers)
%       % \Nnod[1] = element (number of nodes) at index 1
%       \foreachitem \N \in \Nnod{ % loop over layers
%         % \N     = current element in this iteration (i.e. number of nodes for this layer)
%         % \Ncnt  = index of current layer in this iteration
%         \foreach \i [evaluate={\x=\Ncnt; \y=\N/2-\i+0.5; \prev=int(\Ncnt-1);}] in {1,...,\N}{ % loop over nodes
%           \node[mynode] (N\Ncnt-\i) at (\x,\y) {};
%           \ifnum\Ncnt>1 % connect to previous layer
%             \foreach \j in {1,...,\Nnod[\prev]}{ % loop over nodes in previous layer
%               \draw[thick] (N\prev-\j) -- (N\Ncnt-\i); % connect arrows directly
%             }
%           \fi % else: nothing to connect first layer
%         }
%       }
%     \end{tikzpicture}
%     \caption[Neural network architecture for synthetic data]{For synthetic data calibration, the NN architecture is the following: 1D input layer with an ReLU activation function, 10 node hidden layer with a leakyReLU activation function, and 2D output.}
%     \label{nn_synthetic_data}
% \end{figure}

\subsection{Numerical Result}
\begin{figure}[H]
    \centering
    \begin{subfigure}[b]{0.33\textwidth}
        \centering
        \includegraphics[width=\textwidth]{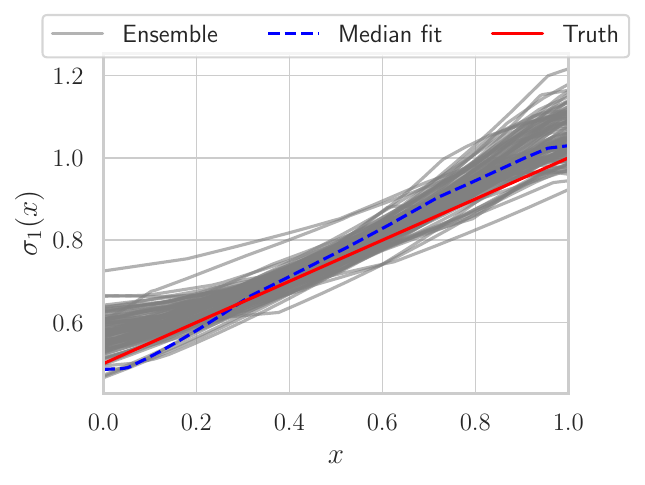}
        \caption{}
    \end{subfigure}
    \begin{subfigure}[b]{0.33\textwidth}  
        \centering 
        \includegraphics[width=\textwidth]{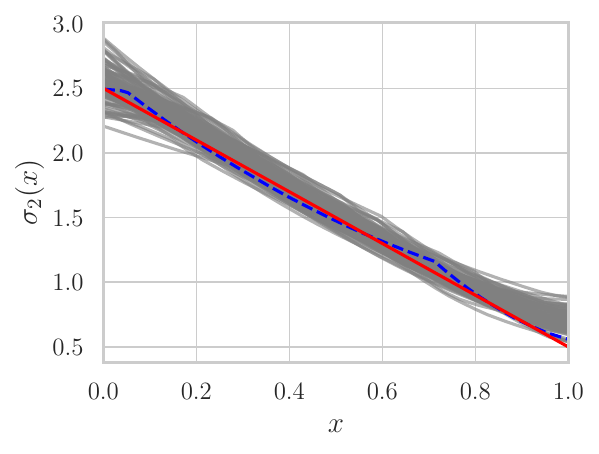}
        \caption{}
    \end{subfigure}
    \begin{subfigure}[b]{0.32\textwidth}   
        \centering 
        \includegraphics[width=\textwidth]{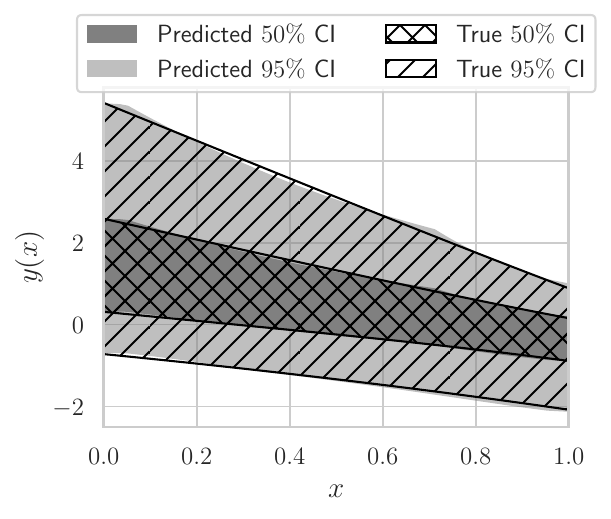}
        \caption{}
        \label{NO_TPG_ci}
    \end{subfigure}
    \vspace{-1\baselineskip}     \caption{Numerical results for scenario A: TPG with linear parameters}
    \label{NO_TPG}
\end{figure}

\begin{figure}[H]
    \centering
    \begin{subfigure}[b]{0.33\textwidth}
        \centering
        \includegraphics[width=\textwidth]{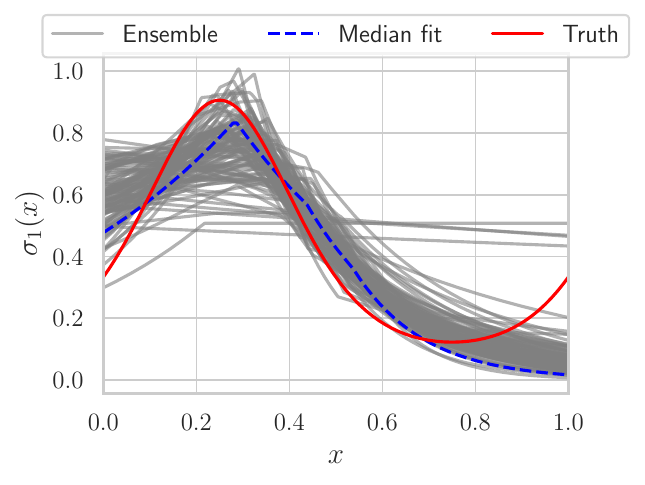}
        \caption{}
    \end{subfigure}
    \begin{subfigure}[b]{0.33\textwidth}  
        \centering 
        \includegraphics[width=\textwidth]{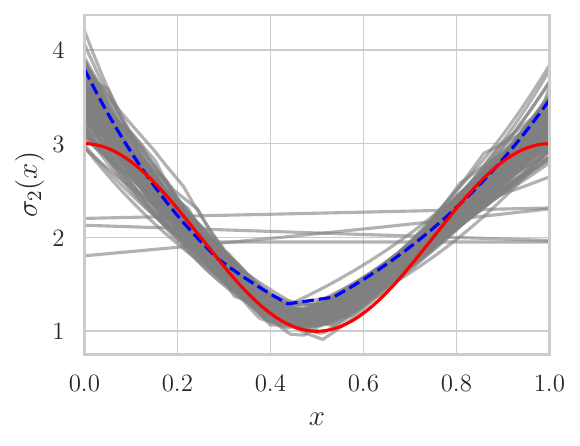}
        \caption{}
    \end{subfigure}
    \begin{subfigure}[b]{0.32\textwidth}   
        \centering 
        \includegraphics[width=\textwidth]{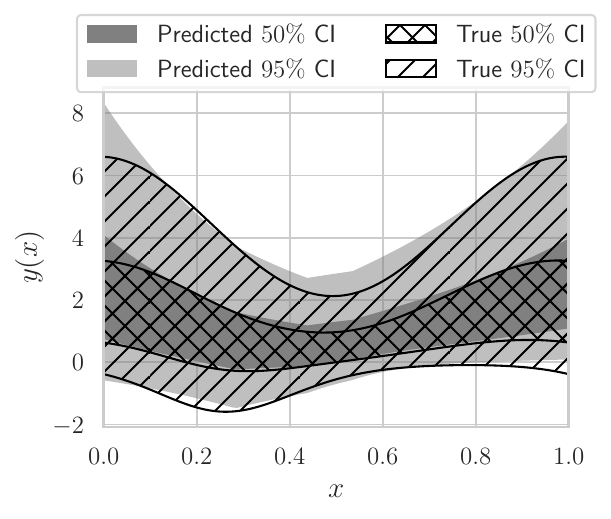}
        \caption{}
        \label{NO_trig_TPG_ci}
    \end{subfigure}
    \vspace{-1\baselineskip}     \caption{Numerical results for scenario B: TPG with trig parameters}
    \label{NO_trig_TPG}
\end{figure}

\begin{figure}[H]
    \centering
    \begin{subfigure}[b]{0.33\textwidth}
        \centering
        \includegraphics[width=\textwidth]{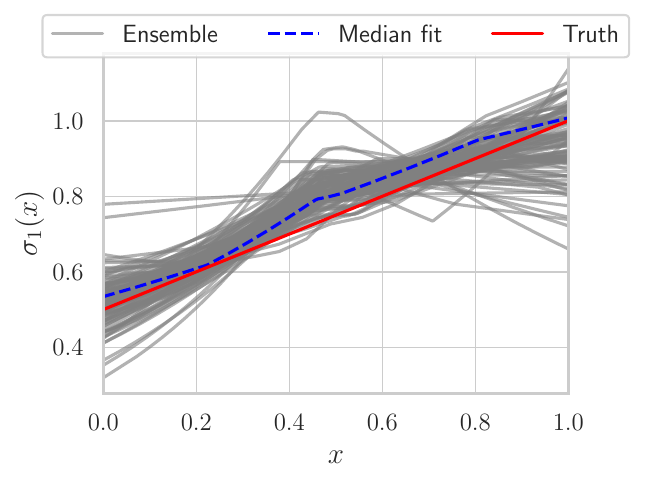}
        \caption{}
    \end{subfigure}
    \begin{subfigure}[b]{0.33\textwidth}  
        \centering 
        \includegraphics[width=\textwidth]{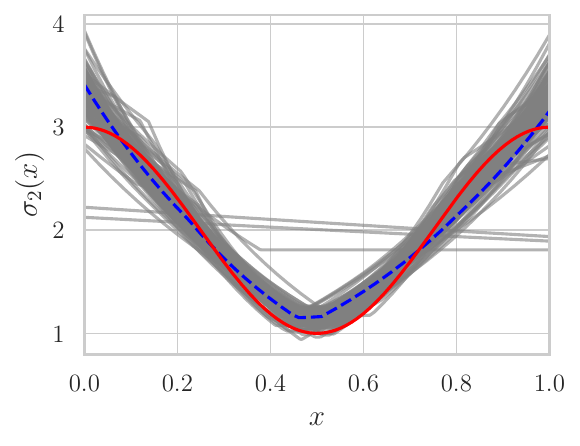}
        \caption{}
    \end{subfigure}
    \begin{subfigure}[b]{0.32\textwidth}   
        \centering 
        \includegraphics[width=\textwidth]{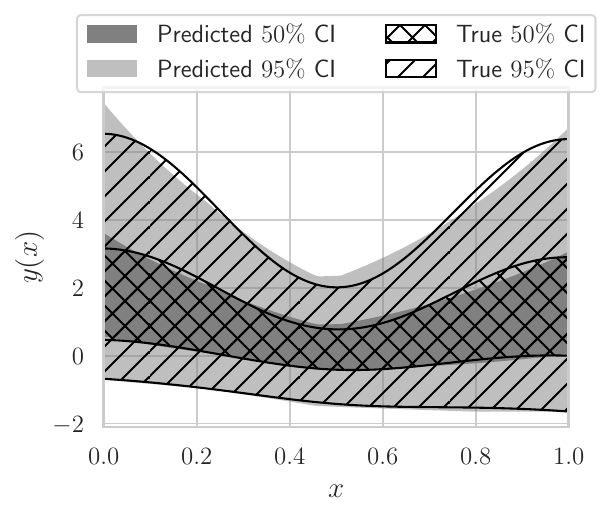}
        \caption{}   
        \label{NO_linear_trig_TPG_ci}
    \end{subfigure}
    \vspace{-1\baselineskip}     \caption{Numerical results for scenario C: TPG with combo parameters}
    \label{NO_linear_trig_TPG}
\end{figure}

\begin{figure}[H]
    \centering
    \begin{subfigure}[b]{0.33\textwidth}
        \centering
        \includegraphics[width=\textwidth]{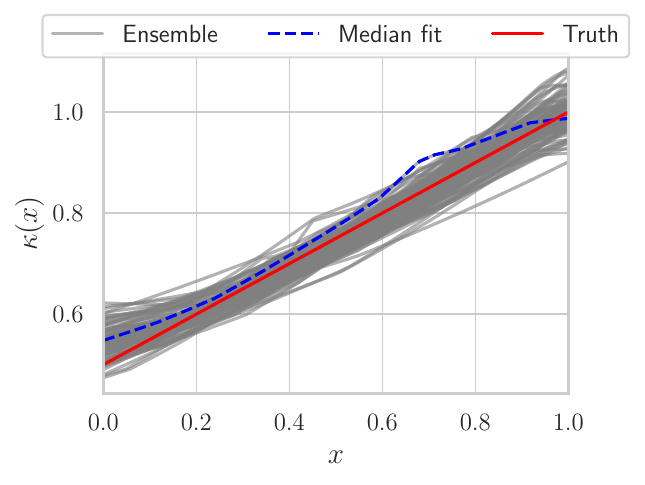}
        \caption{}
    \end{subfigure}
    \begin{subfigure}[b]{0.33\textwidth}  
        \centering 
        \includegraphics[width=\textwidth]{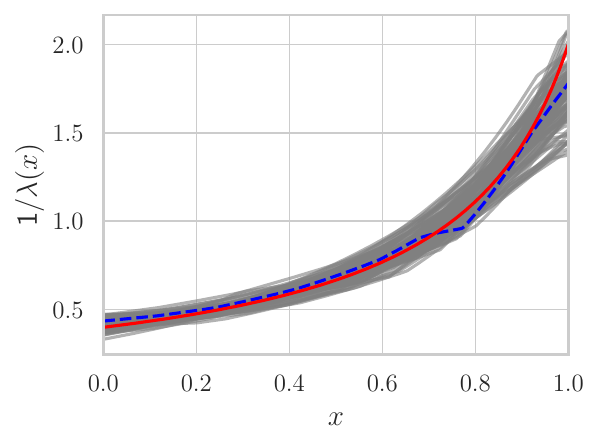}
        \caption{}
    \end{subfigure}
    \begin{subfigure}[b]{0.32\textwidth}   
        \centering 
        \includegraphics[width=\textwidth]{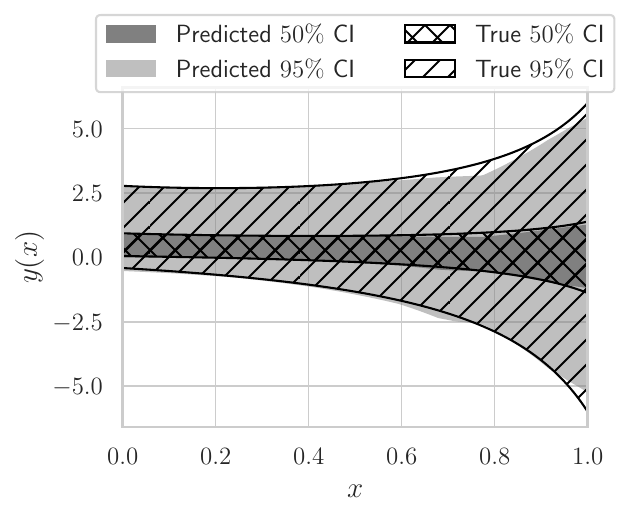}
        \caption{}
        \label{NO_AL_ci}
    \end{subfigure}
    \vspace{-1\baselineskip}     \caption{Numerical results for scenario D: AL with linear parameters}
    \label{NO_AL}
\end{figure}

\begin{figure}[H]
    \centering
    \begin{subfigure}[b]{0.33\textwidth}
        \centering
        \includegraphics[width=\textwidth]{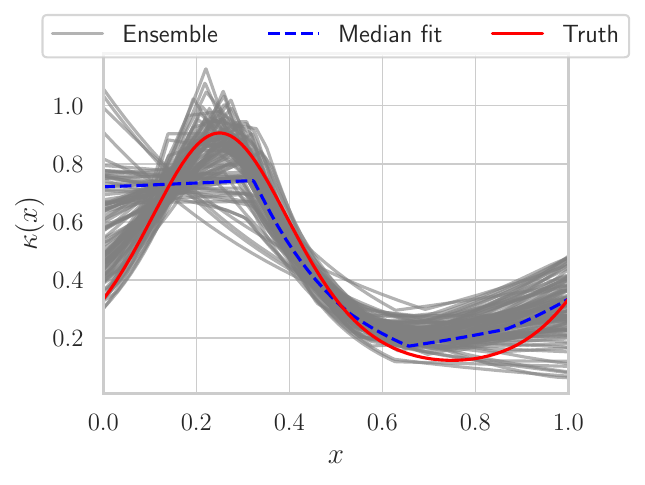}
        \caption{}
    \end{subfigure}
    \begin{subfigure}[b]{0.33\textwidth}  
        \centering 
        \includegraphics[width=\textwidth]{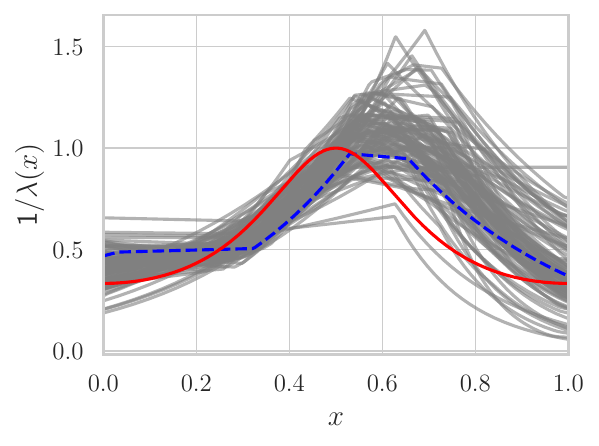}
        \caption{}
    \end{subfigure}
    \begin{subfigure}[b]{0.32\textwidth}   
        \centering 
        \includegraphics[width=\textwidth]{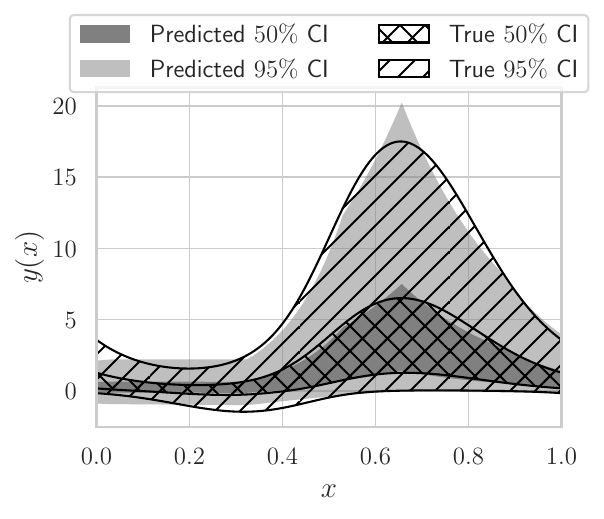}
        \caption{}  
        \label{NO_trig_AL_ci}
    \end{subfigure}
    \vspace{-1\baselineskip}     \caption{Numerical results for scenario E: AL with trig parameters}
    \label{NO_trig_AL}
\end{figure}

\begin{figure}[H]
    \centering
    \begin{subfigure}[b]{0.33\textwidth}
        \centering
        \includegraphics[width=\textwidth]{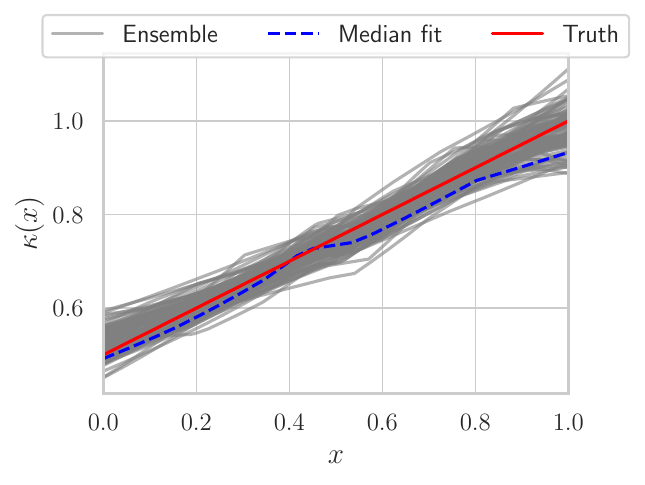}
        \caption{}
    \end{subfigure}
    \begin{subfigure}[b]{0.33\textwidth}  
        \centering 
        \includegraphics[width=\textwidth]{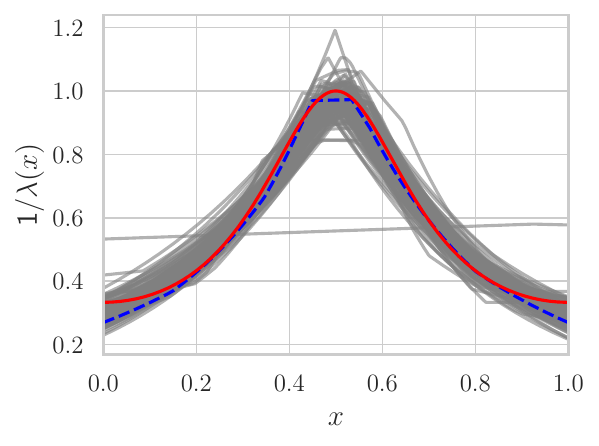}
        \caption{}
    \end{subfigure}
    \begin{subfigure}[b]{0.32\textwidth}   
        \centering 
        \includegraphics[width=\textwidth]{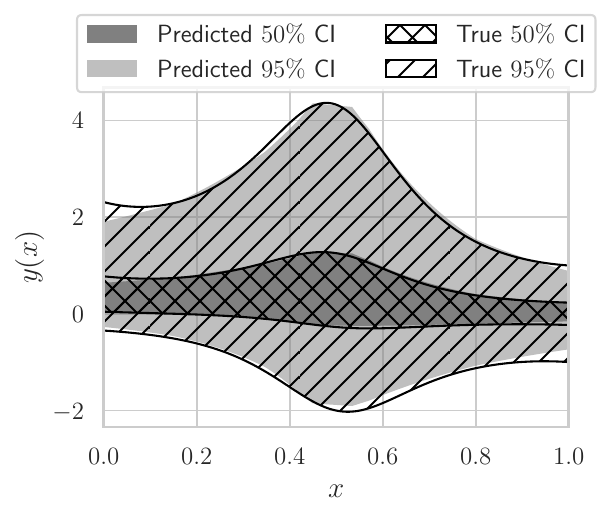}
        \caption{}   
        \label{NO_linear_trig_AL_ci}
    \end{subfigure}
    \vspace{-1\baselineskip}     \caption{Numerical results for scenario F: AL with combo parameters}
    \label{NO_linear_trig_AL}
\end{figure}

Here we present numerical results for the six calibration scenarios in \Crefrange{NO_TPG}{NO_linear_trig_AL}. In subfigures (a) and (b) we compare the true parameter functions of the measurement errors to the learned parameter functions of $100$ NN ensembles. From those ensembles, we select the ensemble member that generates median test loss, where the testing data is a new set of $2,000$ observation-prediction pairs. Then, the median fit ensemble member is used as a representative model for predicting measurement error in the output. In subfigure (c) we compare the true measurement error to the predicted measurement error by plotting their $50\%$ and $95\%$ CIs. The resulting measurement error in the calibration scenarios can be right-skewed, left-skewed, or symmetric about $y(x)=0$. 

Overall, the NN ensembles learn the trend in the true parameter functions. Linear parameter functions are easier for the NNs to learn, but the ensembles agree with nonlinear, trigonometric functions and combinations of linear and nonlinear functions. When comparing the true measurement error to the predicted error from the median NN ensemble member, predicted CIs closely align with the truth. There are very few visual discrepancies in true and predicted $50\%$ CIs. Discrepancies increase slightly for $95\%$ CIs, and we hypothesize these discrepancies result from less data. 

\subsection{Misspecified Measurement Error Distribution}
In the calibration scenarios above, noisy observations were generated from the same distribution we learned. When given real-world data, we often will not know the underlying distribution or the measurement error may not follow a single distribution. Therefore, we investigate a synthetic example with measurement errors learned from misspecified distributions. The true measurement error is generated from a gamma distribution with a PDF of
\begin{equation*}
    p_k(\varepsilon) = \frac{\beta(\bm X_k)^{\alpha(\bm X_k)}}{(\alpha(\bm X_k) -1)!} \varepsilon^{\alpha(\bm X_k)-1}\exp{\left(-\beta(\bm X_k) \varepsilon \right)},
\end{equation*}
where $\alpha(\bm X_k), \beta(\bm X_k) \in \mathbb{R}_{>0}$ are input-dependent shape and rate parameters. Resulting noisy observations are
\begin{equation*}
    y_k = 0 - \Gamma \left( \alpha(x_k), \beta(x_k)\right),
\end{equation*}
and the distribution parameter functions are the trigonometric functions defined in \Cref{accrue_calibration},  $\bm \theta^{\textbf{trig}}(x_k)  = \left(\theta_1^{\textbf{trig}} (x_k), \theta_2^{\textbf{trig}} (x_k) \right) = \left(\alpha(x_k), \beta(x_k) \right)$. After generating observations, we calibrate the measurement error using both two-piece Gaussian and asymmetric Laplace distributions following the same framework as \Cref{accrue_calibration}. Once we have $100$ trained NN ensembles, the median fit NN is chosen using a new testing set of $2,000$ observation-prediction pairs, and the median NN is used to make predictions.

\begin{figure}[H]
    \centering
    \begin{subfigure}[b]{0.33\textwidth}
        \centering
        \includegraphics[width=\textwidth]{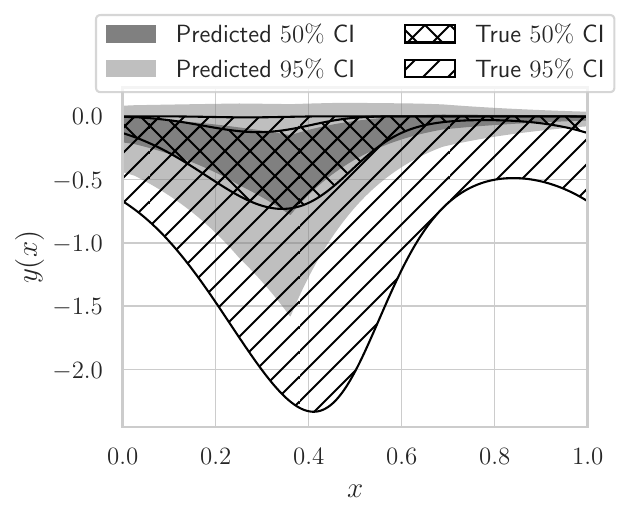}
        \caption{Two-Piece Gaussian}
    \end{subfigure}
    \begin{subfigure}[b]{0.33\textwidth}  
        \centering 
        \includegraphics[width=\textwidth]{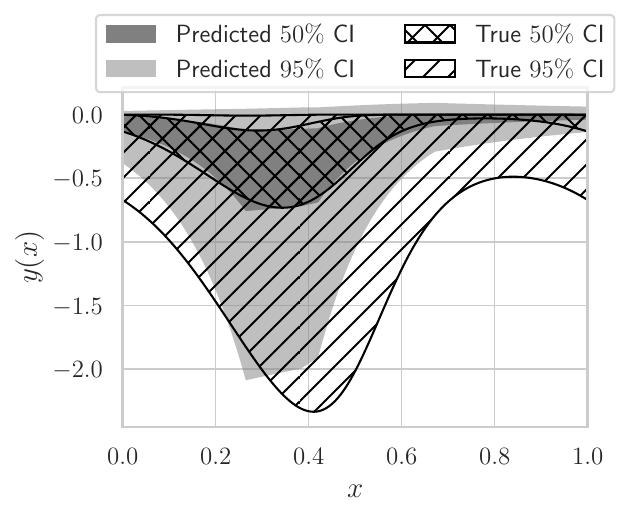}
        \caption{Asymmetric Laplace}
    \end{subfigure}
    \vspace{-1\baselineskip}
    \caption[Numerical results for the misspecified distribution]{Numerical results for the misspecified distribution, where the true measurement error is gamma distributed. Predicted errors are (a) two-piece Gaussian distributed or (b) asymmetric Laplace distributed.}
    \label{Gamma}
\end{figure}

\textbf{Misspecified Numerical Results:} For the misspecified scenario, it is not helpful to compare the true parameter functions to the $100$ NN ensemble predictions since they are input into different distributions. For example, we do not expect a two-piece Gaussian distribution's scale parameters to match a gamma distribution's shape and rate parameters even if the resulting distributions are similar. Instead, in \Cref{Gamma} we only compare outputs: the resulting CIs from the true gamma distribution versus the predicted CIs from the two-piece Gaussian distribution and asymmetric Laplace distribution. The predicted $50\%$ CIs match closely to the truth, but the predicted $95\%$ CIs tend to underestimate. 

Next, we would like to systematically select a prediction distribution (either two-piece Gaussian or asymmetric Laplace). Assuming there is no prior knowledge to inform the decision, we compare the median NN testing losses from the two distributions and select the one with the minimum loss. The predicted distributions from \Cref{Gamma} yielded similar losses shown in \Cref{misspecified_loss}. For this misspecified example, we select the asymmetric Laplace distribution as our prediction distribution. We hypothesize that the asymmetric Laplace distribution is better at modeling the gamma distributed error since both distributions are heavy-tailed.

\begin{table}[h]
\centering
\begin{tabular}{|c|c|c|}
\hline
\textbf{Predicted Distribution} & TPG   & AL    \\ \hline
\textbf{Testing Loss}           & 0.219 & 0.215 \\ \hline
\end{tabular}
\caption{Predicted losses for the misspecified gamma distributed errors}
\label{misspecified_loss}
\end{table}
\section{Weather Forecasting Results}
\label{sec:weather}
In this real-world application, we aim to enhance the reliability of real-time weather forecasts by incorporating uncertainty quantification with ACCRUE and compare against state-of-the-art methods for probabilistic predictions of uncertainty. We focus on temperature forecasting at Denver International Airport (DIA), using data sourced from the National Oceanic and Atmospheric Administration (NOAA) Integrated Surface Database (ISD), which provides high-quality, hourly observations of meteorological variables \cite{smith2011integrated}. For this study, we collected hourly temperature measurements in Celsius at DIA from January 1, 2022 to July 1, 2023. These ground-truth observations serve as the benchmark against which forecast accuracy and uncertainty are evaluated.

More specifically, the model data are one-hour ahead temperature forecasts from the NOAA High-Resolution Rapid Refresh (HRRR) real-time weather forecasting model \cite{james2022high}. While HRRR provides deterministic forecasts, our goal is to estimate the uncertainty around those predictions. To do this, we apply ACCRUE using inputs from the previous hour's dew point, wind speed, and surface pressure—all derived from the same NOAA ISD database. We train ACCRUE on the temperature discrepancies between ISD observations and HRRR one-hour ahead forecasts for a year of data from January 1, 2022 to January 1, 2023 ($8783$ hours). The NN architecture and hyperparameters are the same as for the synthetic results in \Cref{accrue_calibration}, but the input layer is now three-dimensional. Testing data spans January 2, 2023 to July 1, 2023 ($4321$ hours). 

The results in \Cref{weather_loss} show that incorporating uncertainty quantification leads to improved probabilistic forecast performance compared to the deterministic HRRR baseline, as reflected in CRPS-based metrics.
We compare our uncertainty representation using ACCRUE to the performance of deterministic forecasts without uncertainty quantification, as well as two state-of-the-art uncertainty methods: conformal prediction (CP) \cite{vovk2005algorithmic, shafer2008tutorial} and EasyUQ \cite{walz2024easy}. For CP, we follow the Studentized least squares prediction machine formulation \cite[Algorithm 7.2]{vovk2005algorithmic}, using the single-valued model output as the sole covariate. EasyUQ constructs predictive distributions by post-processing model outputs with a learned uncertainty mapping; we use the authors’ released implementation for both methods \cite{walz2024easyuq_github}.
All methods are assessed on held-out testing data.

For the deterministic forecasts generated by the HRRR model, we report the mean absolute error (MAE), which is equivalent to the CRPS in the case of deterministic predictions \cite{hersbach2000decomposition}. For CP and EasyUQ, which output full uncertainty distributions, we compute the mean CRPS. For our approach, we report the ACCRUE loss. ACCRUE achieves the lowest testing loss under its native objective, with the asymmetric Laplace formulation yielding the smallest loss. In terms of mean CRPS, however, all probabilistic methods (CP, EasyUQ, and ACCRUE) perform similarly, with only minor differences across approaches.

% \begin{table}[h]
% \centering
% \resizebox{\textwidth}{!}{%
% \begin{tabular}{|c|c|c|c|c|c|c|}
% \hline
% \multicolumn{2}{|c|}{\textbf{Method}} & HRRR & CP & EasyUQ & TPG & AL \\ \hline
% \multicolumn{2}{|c|}{\textbf{Loss Function}} & MAE & $\overline{\text{CRPS}}$ & $\overline{\text{CRPS}}$ & ACCRUE & ACCRUE \\ \hline
% \multirow{3}{*}{\textbf{Testing Loss}} 
% & 1-hour ahead & 1.1536 & 0.8329 & 0.8305 & 0.7998 & \textbf{0.7495} \\ 
% & 2-hour ahead & x & x & x & x & x \\ 
% & 5-hour ahead & x & x & x & x & x \\ \hline
% \end{tabular}%
% }
% \caption{Predicted losses for temperature forecasts in degrees Celsius.}
% \label{misspecified_loss}
% \end{table}

\begin{table}[ht]
\centering
\resizebox{\textwidth}{!}{%
\begin{tabular}{|c|c|c|c|c|c|c|}
Method & HRRR & CP & EasyUQ & TPG & AL \\ \hline
Loss Function & MAE & $\overline{\text{CRPS}}$ & $\overline{\text{CRPS}}$ & $\overline{\text{CRPS}}$/RS/ACCRUE & $\overline{\text{CRPS}}$/RS/ACCRUE \\ \hline
1-hour Ahead Testing Loss & 1.1536 & 0.8329 & 0.8305 & 0.8324/0.6692/0.7998 & \textbf{0.8312/0.6679/0.7495}
\end{tabular}%
}
\caption{Predicted losses for temperature forecasts in degrees Celsius.}
\label{weather_loss}
\end{table}

\begin{figure}[H]
    \centering
    \begin{subfigure}[b]{0.41\textwidth}
        \centering
        \includegraphics[width=\textwidth]{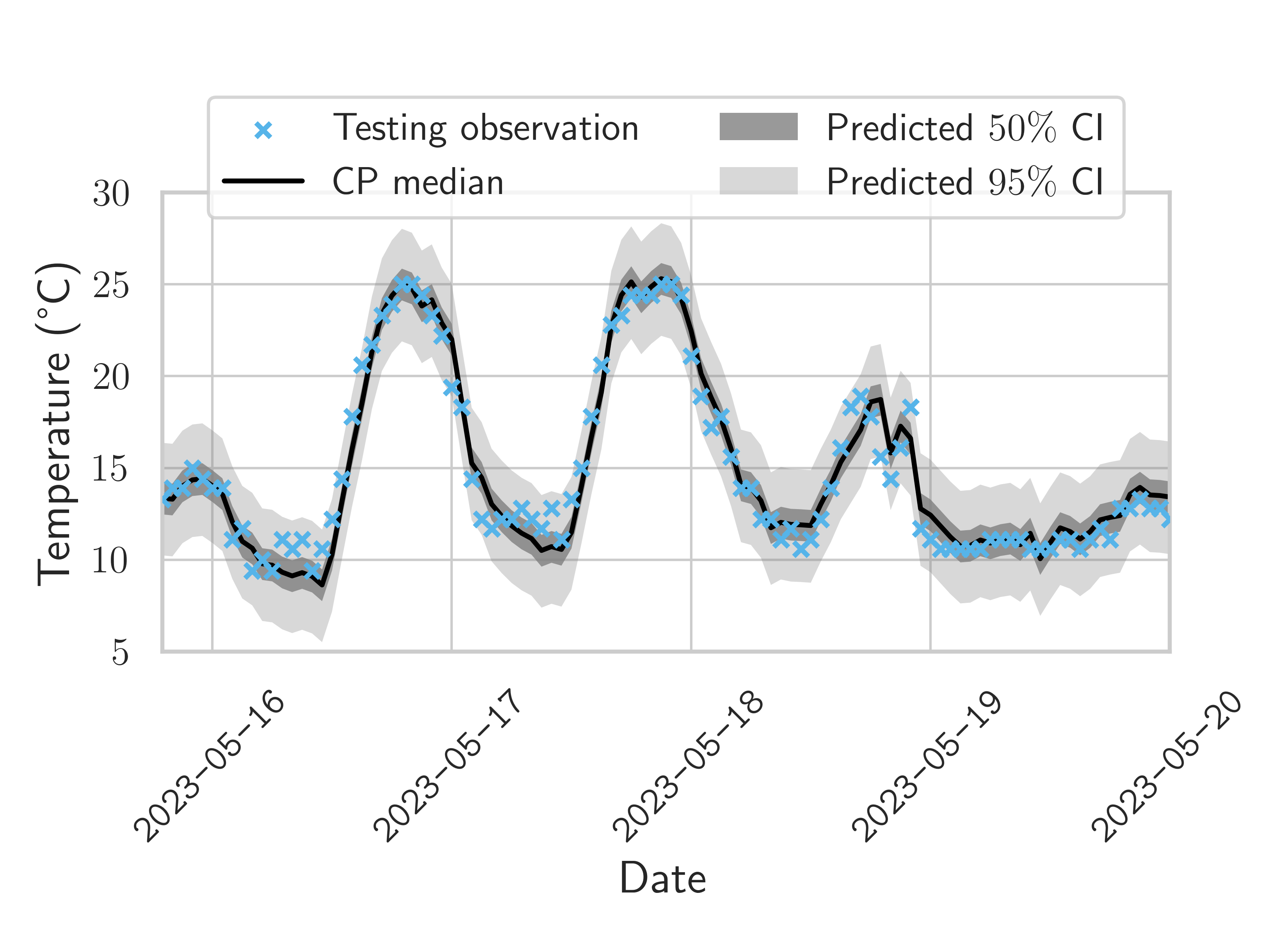}
        \caption{Conformal Prediction}
    \end{subfigure}
    \begin{subfigure}[b]{0.41\textwidth}  
        \centering 
        \includegraphics[width=\textwidth]{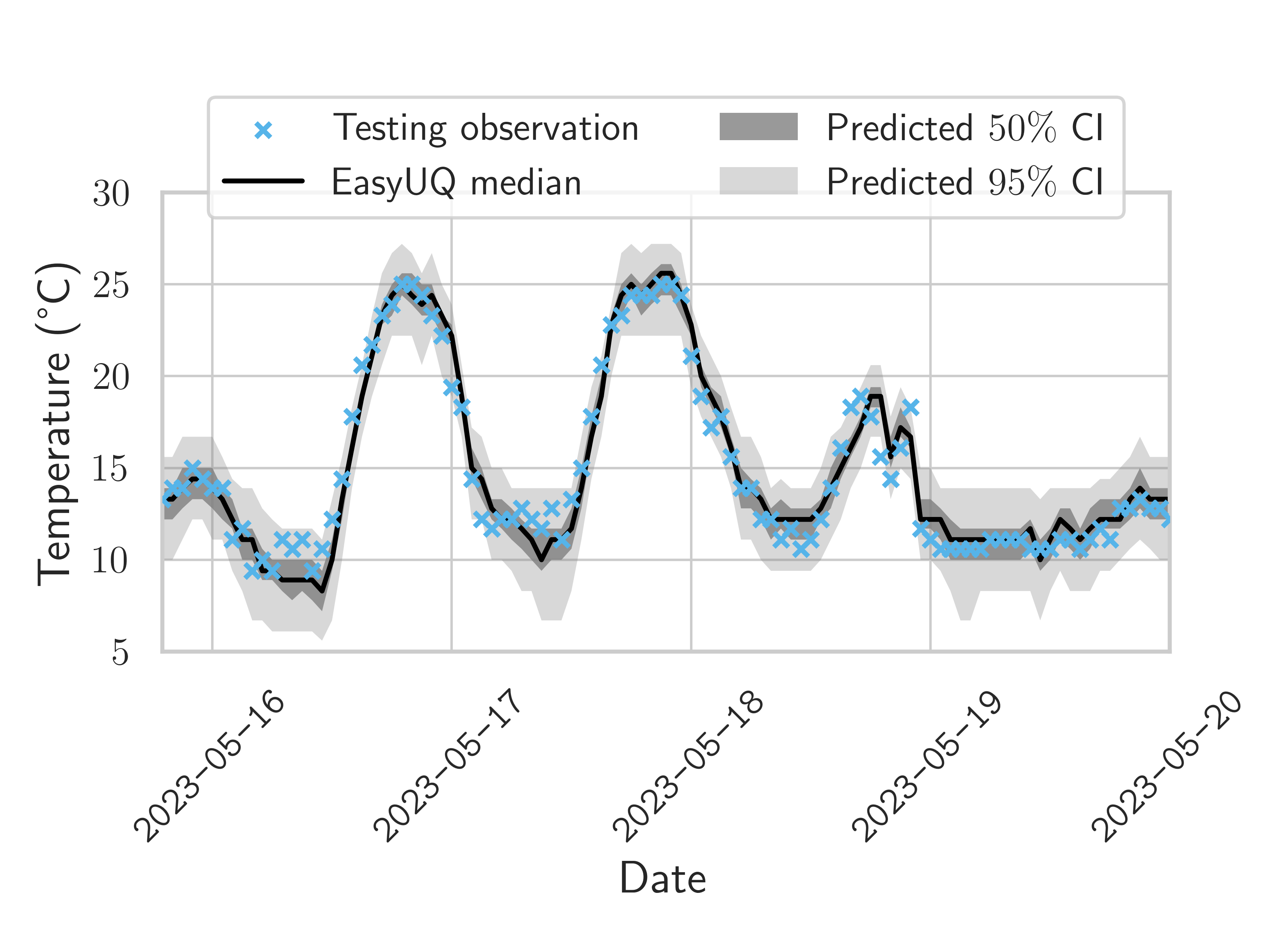}
        \caption{EasyUQ}
    \end{subfigure}
    \par\vspace{-2em}
    \begin{subfigure}[b]{0.41\textwidth}
        \centering
        \includegraphics[width=\textwidth]{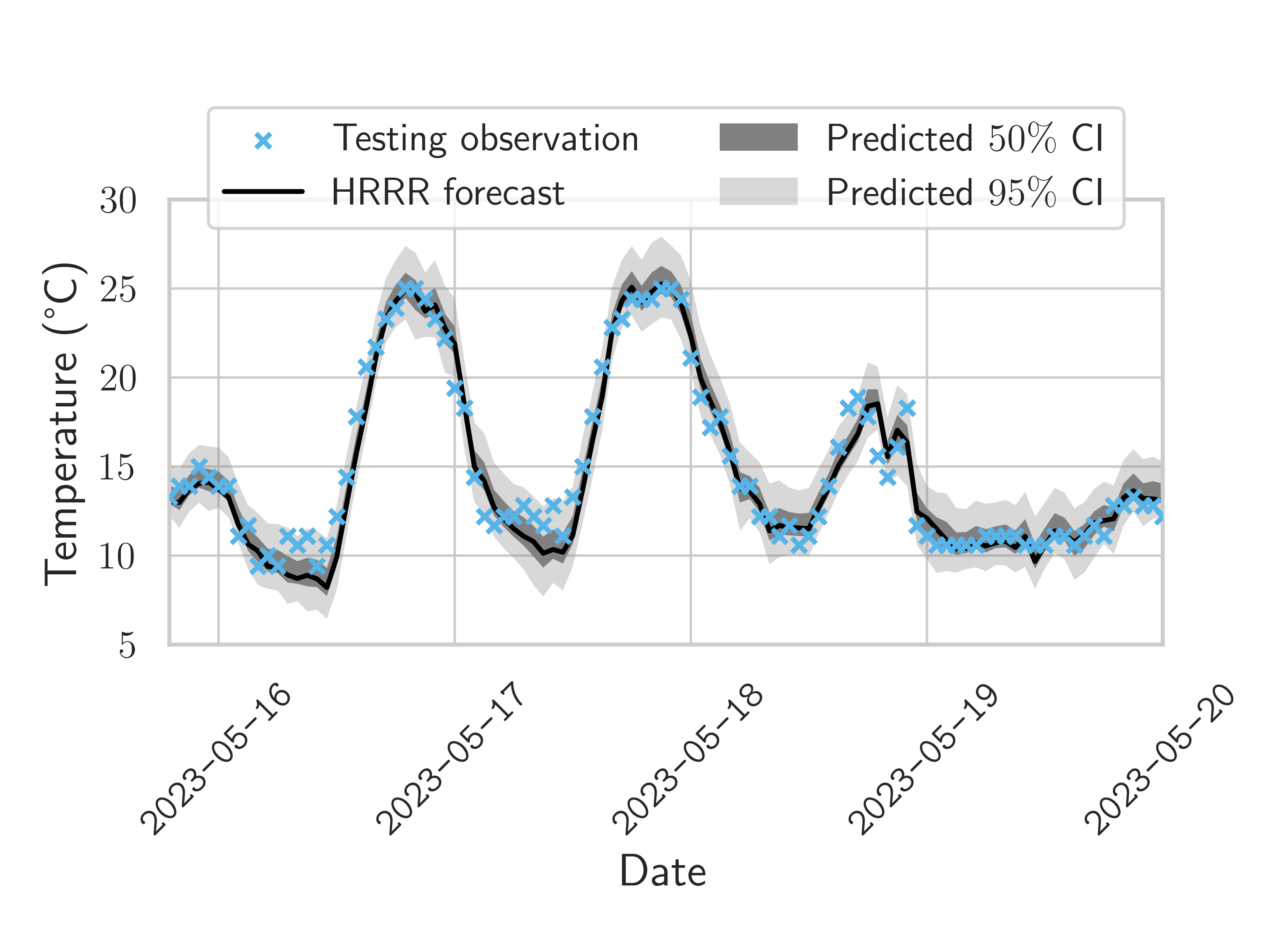}
        \caption{Two-Piece Gaussian}
    \end{subfigure}
    \begin{subfigure}[b]{0.41\textwidth}  
        \centering 
        \includegraphics[width=\textwidth]{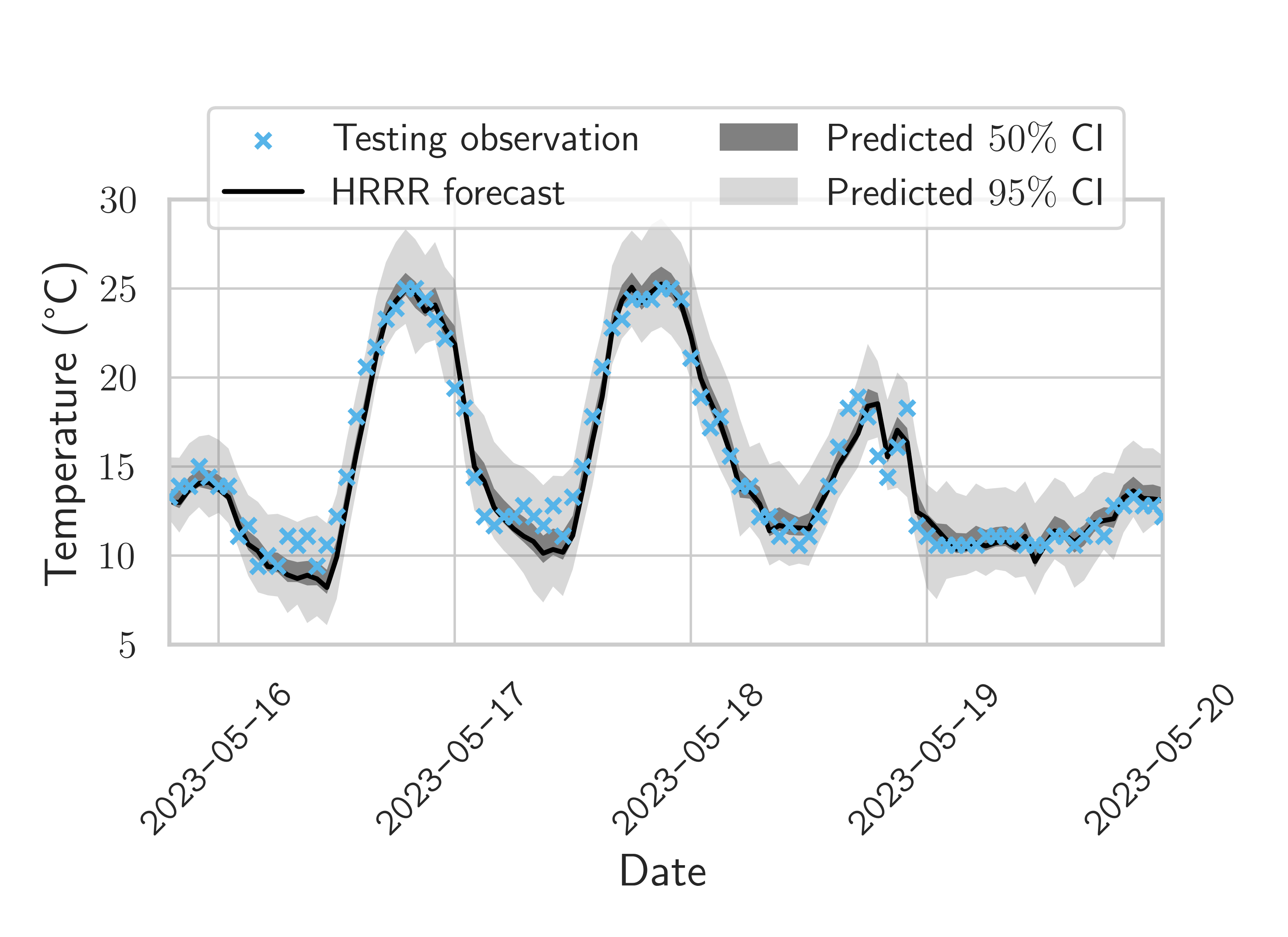}
        \caption{Asymmetric Laplace}
    \end{subfigure}
    \vspace{-1\baselineskip}
    \caption[Numerical results temperature forecasts]{Uncertainty estimates for HRRR one-hour ahead temperature forecasts for testing data extrapolating in time. }
    % Predicted uncertainties are (a) conformal predictions (b) EasyUQ, (c) ACCRUE two-piece Gaussian distributed, or (d) ACCRUE asymmetric Laplace distributed.}
    \label{test_hrrr_times}
\end{figure}

\begin{figure}[H]
    \centering
    \includegraphics[width=1.0\linewidth]{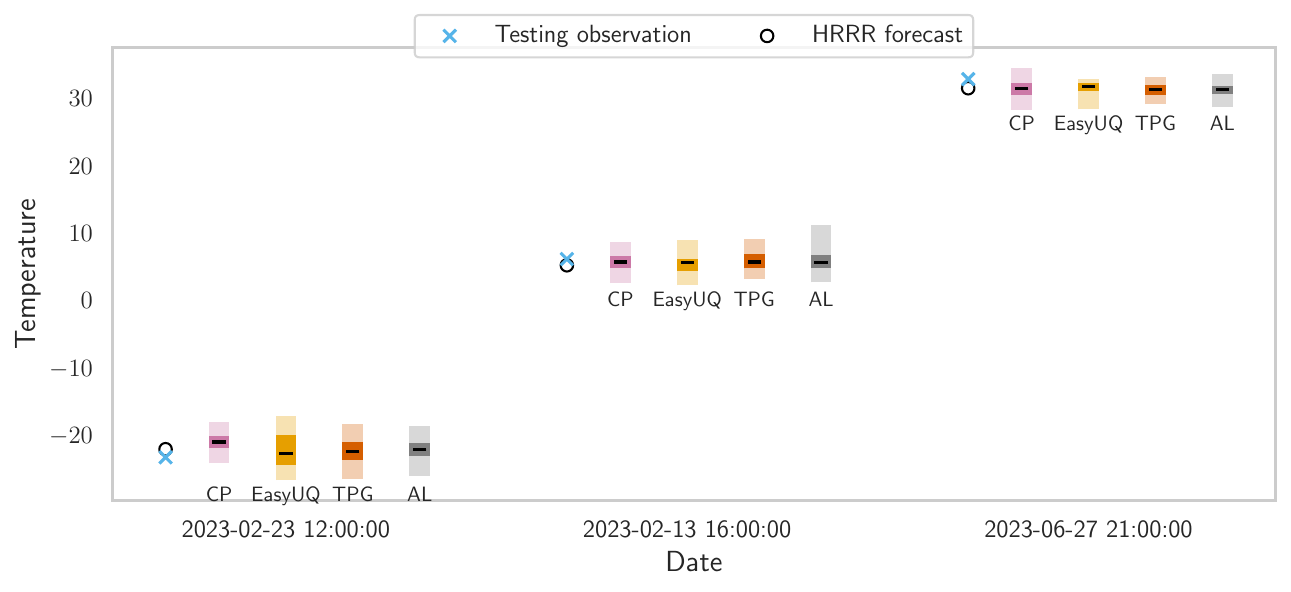}
    \caption{Minimum, median, and maximum testing observations (blue crosses) compared to HRRR One-hour ahead temperature forecasts (black circles), and four uncertainty representations added to the forecasts: conformal prediction (CP), EasyUQ, and ACCRUE with two-piece Gaussian (TPG) and asymmetric Laplace (AL) distributions. Uncertainty representations are shown with $95\%$ and $50\%$ CIs (light and dark shaded regions) and the median (black dash).}
    \label{fig:boxplot1}
\end{figure}

A subset of testing observations from five days in May 2023 is visualized in \Cref{test_hrrr_times}, illustrating how each method represents forecast uncertainty over time.
To further examine these differences, we compare approaches for three representative testing
observations spanning a wide range of temperatures in \Cref{fig:boxplot1}. Across all three cases, the methods produce broadly similar prediction intervals, with comparable spread and alignment relative to the observed temperatures. All approaches capture the observations in each scenario and yield similar uncertainty characterizations, with differences that are relatively minor compared to the overall variability.

\section{Conclusions}
\label{sec:conclusions}
We extend ACCRUE to non-Gaussian uncertainty distributions, specifically two-piece Gaussian and asymmetric Laplace distributions. Then we learn two input-dependent parameter functions that describe the uncertainty at each observation-prediction pair. Synthetic results show we can capture trends of the true parameter functions when the form of the underlying uncertainty distribution is known. When the underlying distribution is known or misspecified, we closely predict the true $50\%$ CIs, but predicted $95\%$ CIs can be less consistent.

In the future, we will investigate more misspecified errors using synthetic and real-world data. We will extend to space weather applications with multidimensional inputs and more complex uncertainties. For example, the severity of geomagnetic storms is often assessed by the disturbance storm time (Dst) index, and many models have been developed to forecast the Dst index \cite{gruet2018multiple, hu2023multi}. Usually, these models are deterministic, and they do not account for uncertainties. However, we want probabilistic forecasts that can address model underpredictions.

% \appendix
% \section{An example appendix} 

\section*{Acknowledgments}
The authors would like to thank Parker Hund and Dhamma Kimpara for many helpful discussions. This work utilized the Alpine high performance computing resource at the University of Colorado Boulder. Alpine is jointly funded by the University of Colorado Boulder, the University of Colorado Anschutz, Colorado State University, and the National Science Foundation (award 2201538). Sandia National Laboratories is a multi-mission laboratory managed and operated by National Technology \& Engineering Solutions of Sandia, LLC (NTESS), a wholly owned subsidiary of Honeywell International Inc., for the U.S. Department of Energy’s National Nuclear Security Administration (DOE/NNSA) under contract DE-NA0003525. This written work is authored by an employee of NTESS. The employee, not NTESS, owns the right, title and interest in and to the written work and is responsible for its contents. Any subjective views or opinions that might be expressed in the written work do not necessarily represent the views of the U.S. Government. The publisher acknowledges that the U.S. Government retains a non-exclusive, paid-up, irrevocable, world-wide license to publish or reproduce the published form of this written work or allow others to do so, for U.S. Government purposes. The DOE will provide public access to results of federally sponsored research in accordance with the DOE Public Access Plan.

\bibliographystyle{siamplain}
\bibliography{references}

@article{camporeale2021accrue,
  title={{ACCRUE}: Accurate and reliable uncertainty estimate in deterministic models},
  author={Camporeale, Enrico and Car{\`e}, Algo},
  journal={International Journal for Uncertainty Quantification},
  volume={11},
  number={4},
  year={2021},
  publisher={Begel House Inc.}
}

@article{camporeale2019generation,
  title={On the generation of probabilistic forecasts from deterministic models},
  author={Camporeale, Enrico and Chu, Xiangning and Agapitov, OV and Bortnik, Jacob},
  journal={Space Weather},
  volume={17},
  number={3},
  pages={455--475},
  year={2019},
  publisher={Wiley Online Library}
}

@article{jordan2017evaluating,
  title={Evaluating probabilistic forecasts with scoring Rules},
  author={Jordan, Alexander and Kr{\"u}ger, Fabian and Lerch, Sebastian},
  journal={arXiv preprint arXiv:1709.04743},
  year={2017}
}

@article{gneiting2010predicting,
  title={Predicting inflation: Professional experts versus no-change forecasts},
  author={Gneiting, Tilmann and Thorarinsdottir, Thordis L},
  journal={arXiv preprint arXiv:1010.2318},
  year={2010}
}

@article{gruet2018multiple,
  title={Multiple-hour-ahead forecast of the {Dst} index using a combination of long short-term memory neural network and Gaussian process},
  author={Gruet, Marina A and Chandorkar, M and Sicard, Ang{\'e}lica and Camporeale, Enrico},
  journal={Space Weather},
  volume={16},
  number={11},
  pages={1882--1896},
  year={2018},
  publisher={Wiley Online Library}
}

@article{hu2023multi,
  title={Multi-Hour-Ahead {Dst} Index Prediction Using Multi-Fidelity Boosted Neural Networks},
  author={Hu, Andong and Camporeale, Enrico and Swiger, Brian},
  journal={Space Weather},
  volume={21},
  number={4},
  pages={e2022SW003286},
  year={2023},
  publisher={Wiley Online Library}
}

@article{eberhard1999automatic,
  title={Automatic differentiation of numerical integration algorithms},
  author={Eberhard, Peter and Bischof, Christian},
  journal={Mathematics of Computation},
  volume={68},
  number={226},
  pages={717--731},
  year={1999}
}

@article{hersbach2000decomposition,
  title={Decomposition of the continuous ranked probability score for ensemble prediction systems},
  author={Hersbach, Hans},
  journal={Weather and Forecasting},
  volume={15},
  number={5},
  pages={559--570},
  year={2000},
  publisher={American Meteorological Society}
}

@article{issan2023bayesian,
  title={Bayesian inference and global sensitivity analysis for ambient solar wind prediction},
  author={Issan, Opal and Riley, Pete and Camporeale, Enrico and Kramer, Boris},
  journal={Space Weather},
  volume={21},
  number={9},
  pages={e2023SW003555},
  year={2023},
  publisher={Wiley Online Library}
}

@article{sagi2018ensemble,
  title={Ensemble learning: A survey},
  author={Sagi, Omer and Rokach, Lior},
  journal={Wiley Interdisciplinary Reviews: Data Mining and Knowledge discovery},
  volume={8},
  number={4},
  pages={e1249},
  year={2018},
  publisher={Wiley Online Library}
}

@article{luengo2020survey,
  title={A survey of {Monte} {Carlo} methods for parameter estimation},
  author={Luengo, David and Martino, Luca and Bugallo, M{\'o}nica and Elvira, V{\'\i}ctor and S{\"a}rkk{\"a}, Simo},
  journal={European Association for Signal Processing Journal on Advances in Signal Processing},
  volume={2020},
  pages={1--62},
  year={2020},
  publisher={Springer}
}

@article{walz2024easy,
  title={Easy Uncertainty Quantification ({EasyUQ}): Generating predictive distributions from single-valued model output},
  author={Walz, Eva-Maria and Henzi, Alexander and Ziegel, Johanna and Gneiting, Tilmann},
  journal={SIAM Review},
  volume={66},
  number={1},
  pages={91--122},
  year={2024},
  publisher={Society for Industrial and Applied Mathematics}
}

@article{taquet2022mapie,
  title={MAPIE: an open-source library for distribution-free uncertainty quantification},
  author={Taquet, Vianney and Blot, Vincent and Morzadec, Thomas and Lacombe, Louis and Brunel, Nicolas},
  journal={arXiv preprint arXiv:2207.12274},
  year={2022}
}

@article{li2013optimal,
  title={Optimal bandwidth selection for nonparametric conditional distribution and quantile functions},
  author={Li, Qi and Lin, Juan and Racine, Jeffrey S},
  journal={Journal of Business \& Economic Statistics},
  volume={31},
  number={1},
  pages={57--65},
  year={2013},
  publisher={Taylor \& Francis}
}

@book{kotu2018data,
  title={Data Science: Concepts and Practice},
  author={Kotu, Vijay and Deshpande, Bala},
  year={2018},
  publisher={Morgan Kaufmann}
}

@article{smith2011integrated,
  title={The integrated surface database: Recent developments and partnerships},
  author={Smith, Adam and Lott, Neal and Vose, Russ},
  journal={Bulletin of the American Meteorological Society},
  volume={92},
  number={6},
  pages={704--708},
  year={2011},
  publisher={JSTOR}
}

@article{james2022high,
  title={The High-Resolution Rapid Refresh (HRRR): an hourly updating convection-allowing forecast model. Part II: Forecast performance},
  author={James, Eric P and Alexander, Curtis R and Dowell, David C and Weygandt, Stephen S and Benjamin, Stanley G and Manikin, Geoffrey S and Brown, John M and Olson, Joseph B and Hu, Ming and Smirnova, Tatiana G and others},
  journal={Weather and Forecasting},
  volume={37},
  number={8},
  pages={1397--1417},
  year={2022}
}

@book{vovk2005algorithmic,
  title={Algorithmic learning in a random world},
  author={Vovk, Vladimir and Gammerman, Alexander and Shafer, Glenn},
  volume={29},
  year={2005},
  publisher={Springer}
}

@article{shafer2008tutorial,
  title={A tutorial on conformal prediction.},
  author={Shafer, Glenn and Vovk, Vladimir},
  journal={Journal of Machine Learning Research},
  volume={9},
  number={3},
  year={2008}
}

@misc{walz2024easyuq_github,
  author       = {Walz, Elias V. and others},
  title        = {EasyUQ: Easy-to-Use Uncertainty Quantification},
  year         = {2024},
  howpublished = {\url{https://github.com/evwalz/easyuq}},
  note         = {GitHub repository}
}

@software{rileigh_bandy_2026_19343211,
  author       = {Rileigh Bandy},
  title        = {rbandy/{ACCRUE}\_for\_skewed: {ACCRUE} for skewed
                   uncertainty distributions
                  },
  month        = mar,
  year         = 2026,
  publisher    = {Zenodo},
  version      = {init},
  doi          = {10.5281/zenodo.19343211},
}

@article{chada2025bayesian,
  title={Bayesian Deep Learning with Multilevel Trace-Class Neural Networks},
  author={Chada, Neil K and Jasra, Ajay and Law, Kody JH and Singh, Sumeetpal S},
  journal={SIAM Journal on Mathematics of Data Science},
  volume={7},
  number={3},
  pages={1210--1240},
  year={2025},
  publisher={SIAM}
}

@article{martin2025online,
  title={Online Machine Teaching under Learner Uncertainty: Gradient Descent Learners of a Quadratic Loss},
  author={Martin-Urcelay, Belen and Rozell, Christopher J and Bloch, Matthieu R},
  journal={SIAM Journal on Mathematics of Data Science},
  volume={7},
  number={3},
  pages={884--905},
  year={2025},
  publisher={SIAM}
}

@article{xie2021nonparametric,
  title={A nonparametric Bayesian framework for uncertainty quantification in stochastic simulation},
  author={Xie, Wei and Li, Cheng and Wu, Yuefeng and Zhang, Pu},
  journal={SIAM/ASA Journal on Uncertainty Quantification},
  volume={9},
  number={4},
  pages={1527--1552},
  year={2021},
  publisher={SIAM}
}

@article{morrison2018representing,
  title={Representing model inadequacy: A stochastic operator approach},
  author={Morrison, Rebecca E and Oliver, Todd A and Moser, Robert D},
  journal={SIAM/ASA Journal on Uncertainty Quantification},
  volume={6},
  number={2},
  pages={457--496},
  year={2018},
  publisher={SIAM}
}

@article{portone2022bayesian,
  title={Bayesian inference of an uncertain generalized diffusion operator},
  author={Portone, Teresa and Moser, Robert D},
  journal={SIAM/ASA Journal on Uncertainty Quantification},
  volume={10},
  number={1},
  pages={151--178},
  year={2022},
  publisher={SIAM}
}

\end{document}